\documentclass[a4paper, 11pt]{article}

\usepackage{jheppub}
\usepackage{orcidlink}

\usepackage[skip=.25\baselineskip]{parskip}
\usepackage{indentfirst}
\usepackage[dvipsnames]{xcolor}
\usepackage{enumitem}
\usepackage[nottoc]{tocbibind}
\usepackage[T1]{fontenc}
\usepackage{lmodern}

\usepackage{hyperref}
\hypersetup{
	colorlinks=true,
	urlcolor=BrickRed,
	linkcolor=OliveGreen,
	citecolor=BrickRed,
	pdftitle={Hawking Amplitudes: Subtle Points About Saddle Points},
	pdfauthor={Owen Lear, Daniel McLoughlin, Sebastian Mizera, John Staunton},
	pdfdisplaydoctitle=true,
	pdfstartview=FitH
}

\numberwithin{equation}{section}
\def\equationautorefname~#1\null{Eq.~(#1)\null}

\makeatletter
\renewcommand*{\eqref}[1]{\hyperref[{#1}]{\textup{\tagform@{\ref*{#1}}}}}
\makeatother

\usepackage{amsmath, amssymb}
\usepackage{physics}
\usepackage{dsfont}
\usepackage[mathscr]{eucal}

\usepackage{graphicx}
\usepackage{tikz}
\usepackage{pgfplots}

\usetikzlibrary{
    arrows.meta,             % For customizable arrow tips
    decorations.pathmorphing,  % For the zigzag cutoff line
    decorations.pathreplacing, % For the curly braces
    positioning,             % For easy node placement
    calc,                     % For coordinate calculations
    patterns,
    matrix, 
    angles,                 % For angles
    quotes,                  % Can provide string commands tikz
intersections, shapes.arrows, shapes.misc, angles, fadings, tikzmark
}

\tikzset{cross/.style={cross out, draw=black, minimum size=2*(#1-\pgflinewidth), inner sep=0pt, outer sep=0pt}, cross/.default={1pt}}

\pgfplotsset{compat=newest}

\def \be {\begin{equation}}
\def \ee {\end{equation}}

\def \geq {\geqslant}

\def \e {\mathrm{e}}

\title{Hawking Amplitudes:\\ Subtle Points About Saddle Points}

\author[1,2]{Owen Lear \orcidlink{0009-0004-6180-9660},}
\author[1,3]{Daniel McLoughlin \orcidlink{0009-0005-3535-2334},}
\author[1]{Sebastian Mizera \orcidlink{0000-0002-8066-5891},}
\author[1,4]{John Staunton \orcidlink{0009-0004-1661-9577}}

\affiliation[1]{
Department of Physics, Columbia University, 538 West 120th Street, New York, NY 10027, USA}
\affiliation[2]{Leinweber Institute for Theoretical Physics, Randall Laboratory of Physics,\\ University of Michigan, 450 Church St, Ann Arbor, MI 48109-1040, USA}
\affiliation[3]{Mani L. Bhaumik Institute for Theoretical Physics, Department of Physics and Astronomy,\\ University of California, Los Angeles, CA 90095, USA}
\affiliation[4]{Center for Particle Cosmology, Department of Physics and Astronomy,\\ University of Pennsylvania, 209 South 33rd St, Philadelphia, PA 19104, USA}

\emailAdd{owenlear@umich.edu}
\emailAdd{dmcloughlin@physics.ucla.edu}
\emailAdd{sebastian.mizera@columbia.edu}
\emailAdd{jstaunto@sas.upenn.edu}

\abstract{
Hawking amplitudes provide a new way of determining the Hawking spectrum of black holes, using on-shell techniques borrowed from the literature on scattering amplitudes. In this paper, we investigate the interplay between the semi-classical and weak-field approximations previously used to compute Hawking amplitudes. We find that these limits do not necessarily commute, highlighting subtleties with the saddle point analyses governing the semi-classical limit. Resolving these issues allows us to compute semi-classical Hawking amplitudes in a range of black-hole backgrounds, including Tangherlini, Reissner--Nordstr\"om, and Hernquist metrics. Moreover, we describe the saddle-point contributions dominating Hawking amplitudes for different energies and propose a new method for determining the temperature of the resulting thermal spectrum based on Stokes lines of Hawking amplitudes.
}

\begin{document}

\setcounter{tocdepth}{2}
\maketitle
\setcounter{page}{2}

\section{Introduction and Literature Review}
\textbf{Background.} One of the most fascinating consequences of quantum field theory in curved spacetime is that the definition of the vacuum is not invariant. As a result, an accelerating observer sees a thermal distribution of particles where a static observer sees none \cite{Davies:1975, Unruh:1976db}, a vacuum state that undergoes cosmological expansion results in the pair production of particles \cite{Parker:1968mv, Parker:1969au}, and an observer in a black hole background sees a thermal distribution of particles \cite{Hawking:1974rv}. Furthermore, it was shown in recent years that different observers in a black hole background (e.g. static and free-falling) see different temperatures \cite{Barcelo:2010xk, Barcelo:2010pj, Barbado:2011dx, McMaken:2024}.

For the case of black holes, there is an endless number of methods for computing the temperature measured by an observer. One method is to compute the \textit{Bogoliubov coefficients}, which quantify the relationship between two different observers' vacua and therefore encode the number of particles seen by one observer in the others' vacuum. In the context of a collapsing black hole, these Bogoliubov coefficients depend on the null geodesics that start from past null infinity, before the collapse, and reach future null infinity after the collapse. Indeed, this is the method used in Hawking's original paper on particle production in a black hole background \cite{Hawking:1974rv}. A second method involves computing the overlap of vacua directly from the path integral, which additionally allows one to study the increase in entropy resulting from one region of spacetime becoming causally disconnected \cite{Harlow:2014yka, Padmanabhan:2016xjk}. A third method leverages the fact that black hole spacetimes can be mapped to polar-like coordinates, where all continuous functions, including the thermal Green's function, are periodic with period $P$. Since the thermal Green's function is periodic in the inverse temperature, $\beta$, the temperature is $1/P$. A yet different method uses a WKB approximation, which was shown by \cite{Parikh:1999mf} to allow for the inclusion of backreaction and therefore non-thermal contributions to the spectrum. 

On the other hand, there has been a lot of recent attention paid to reformulating this physics in terms of modern field theoretic methods, including on-shell amplitudes.
In \cite{Goldberger:2020geb, Goldberger:2020wbx}, Goldberger and Rothstein computed the Hawking spectrum with the use of the worldline effective field theory by computing the scattering of a scalar field off of a black hole background exchanging off-shell gravitons that are emitted or absorbed by the horizon. In \cite{Aoude:2024sve}, Aoude, O'Connell, and Sergola re-derived the Hawking spectrum using the Kosower--Maybee--O'Connell (KMOC) \cite{Kosower:2018adc} formalism, which constructs in-in observables from on-shell scattering amplitudes. Since then, a double copy relationship between the Hawking spectrum and QCD distributions was noted in \cite{Aoude:2025jvt, Ilderton:2025aql, Carrasco:2025bgu}. Additional work demonstrated that the absorption and emission thermal spectra can be derived by computing a scattering matrix \cite{Aoki:2025ihc} or by bootstrapping the three-particle on-shell amplitude of a massless particle and two unequal mass spinning states and enforcing that the \textit{macroscopic} average of those spinning states coincides with the value of the classical spin of the black hole \cite{Akpinar:2026oni}. In each case, the method works for a static observer far away from a four-dimensional Schwarzschild black hole.

\textbf{Motivation.} On-shell \cite{Kosower:2018adc} and effective field theory \cite{Goldberger:2004jt} methods have been used extensively to successfully compute state-of-the-art black hole observables, including but not limited to compact object observables \cite{Caron-Huot:2025tlq, Ivanov:2026icp, Pani:2025qxs, Solon:2026ubm, Apostolidis:2026qsg}, gravitational waveforms \cite{Cristofoli:2021vyo, Elkhidir:2023dco, Brandhuber:2023hhy, Herderschee:2023fxh, Georgoudis:2023lgf, Caron-Huot:2023vxl, Chang:2026eti}, and impulses \cite{Bern:2021dqo, Herrmann:2021tct, Driesse:2026qiz}. This progress has therefore raised the question whether \textit{quantum} effects in curved backgrounds can also be systematically computed using the same methods. In that direction, techniques developed to study the analytic properties of scattering amplitudes have been applied to Bogoliubov coefficients, such as to rigorously demonstrate a crossing relationship between them in \cite{Aoude:2026bek}. Similarly, scattering amplitude techniques have been used to compute particle production rates due to cosmological expansion \cite{Komissarov:2022gax, Aoki:2024bpj}. By starting from very general considerations and systematically incorporating effects due to finite size, spin, and more, both methods have proven to be very flexible. It is therefore worthwhile to explore the application of on-shell scattering amplitudes techniques to more general situations, such as higher-dimensional and charged black holes. 

However, the extraction of Hawking radiation from amplitudes is more difficult compared to traditional methods.
As we will show in this paper, for black holes other than four-dimensional Schwarzschild one would in principle require expanding to all orders in perturbation theory to extract the correct temperature. Fundamentally, this issue is tied to the interplay between the geometric optics (eikonal or semi-classical) and weak-field (post-Minkowski) approximations that are employed in on-shell techniques.

If we call the characteristic size of the black hole $r_s$, the momentum transfer between the black hole and the probe (massless) particle $q$, and the energy of the particle $E$, then two approximations can be stated as:
\begin{align}
\text{Geometric optics approximation:} &\qquad \abs{q} / E \;\ll 1\, , \\
\text{Weak-field approximation:} &\qquad r_s \abs{q} \;\;\,\ll 1\, .
\end{align}
Our primary goal in this paper is to explore the relationship between these two limits, in particular demonstrating that they do not commute. The way in which we implement the geometric optics expansion in this paper is not new; for similar treatments as applied to the problem of Hawking radiation, see \cite{Srinivasan:1998ty, Parikh:1999mf, Ilderton:2025umd}. Subtleties having to do with the semi-classical limit, in the context of gravitational scattering, were previously discussed in \cite{Cristofoli:2021jas, Bellazzini:2022wzv}.

\textbf{Summary of the Results.} In this paper, we approach the scattering amplitudes calculation by first resumming all diagrams that appear in a weak-field expansion and then using a geometric optics approximation on the resulting amplitude. We then propose a novel method for extracting the temperature of the resulting thermal distribution. Our first main result is to demonstrate that the eikonal and weak field limits do not commute. We use higher-dimensional neutral and charged black holes, called Schwarzschild--Tangherlini and Reissner--Nordstr\"{o}m respectively, as worked examples. We include additional examples in appendices, including for a black hole surrounded by a Hernquist dark matter cloud and for a static observer a finite distance away from the black hole. Our second result, therefore, is to apply the setup of \cite{Aoude:2024sve} to a wider class of black holes. 

Physically, the non-commutativity of each limit is closely related to the fact that the most relevant null rays at infinity are the ones that were near the black hole horizon at the time of collapse. To see this, we note that the geometric optics approximation, characterized by $\eta = \abs{q}/E$, is related to the weak-field approximation, characterized by $h \sim r_s \abs{q}$, by $h/\eta \sim r_s E$, where $h$ will more concretely be identified as the Kerr--Schild potential in the next section. However, the rays we are considering have wavelengths much smaller than the characteristic size of the black hole, $r_s$. Hence, for the Hawking radiation calculation, we must have $h/\eta \gg 1$. This would be incompatible with taking the weak-field approximation before the eikonal approximation. 

In these calculations, the scattering amplitudes calculation is used to compute the Bogoliubov coefficients $\alpha$ and $\beta$. Once one has the Bogoliubov coefficients, the traditional way of computing the Hawking spectrum is to relate $\alpha$ and $\beta$ and then use the unitarity relationship $\abs{\alpha}^2 - \abs{\beta}^2 = 1$ to isolate $\abs{\beta}^2$, which encodes the number of particles at a given energy. In this paper, our third main result is to instead propose a novel way of extracting the temperature from the Bogoliubov coefficient. In particular, we demonstrate that in all of the studied cases, there is a Stokes line, where the asymptotic expansion of the Bogoliubov coefficient discontinuously changes, in the complex plane of the outgoing energy, $E$, located at $\text{Im}\left[E\right] = 2\pi T_H$. Therefore, the computation of the temperature reduces to studying the large energy-behavior of the Bogoliubov coefficient, without needing to solve any integrals. The presence of a Stokes line is often associated with non-perturbative physics and we discuss how this connects the geometric optics and weak-field expansions. 

\textbf{Relation to Previous Work.} Ref.~\cite{Aoude:2024sve}, re-derives the Hawking spectrum through a KMOC-like set-up. By considering the scattering of infalling radiation in a Vaidya collapse background, they are able to compute a \textit{Hawking Amplitude}, $\mathcal{A}$, as well as show how it is related to the Bogoliubov coefficient, $\alpha$. The authors compute this amplitude at tree-level and then demonstrate that in the geometric optics approximation, the tree-level result exponentiates, just as happens in gravitational wave problems \cite{DiVecchia:2023}. They were then able to recover the correct thermal distribution for the four-dimensional Schwarzschild black hole. In our work, we take a similar approach, but attempt to compute the Bogoliubov coefficients by only using eikonal approximations, without additional post-Minkowski expansions. We then apply this strategy to a wider class of black hole solutions.    

One of the examples we study involves a static observer in a Schwarzschild background. Using the techniques of \cite{Barcelo:2010xk, Barcelo:2010pj}, Ref.~\cite{Barbado:2011dx} derives the temperature as seen by both a static observer and free-falling observer. This strategy was later extended for different observers in a Kerr background \cite{McMaken:2024}. Ref.~\cite{Barcelo:2010xk} employs a saddle-point approximation on the Bogoliubov coefficients and shows that they have a Boltzmann distribution. Upon using the unitarity constraint, $\abs{\alpha}^2 - \abs{\beta}^2 = 1$, one then shows the Hawking spectrum is equivalent to a Planck spectrum. In our paper, we further explore the saddle-point approximation on the Bogoliubov coefficients by \textit{complexifying} the outgoing energy $E$. In doing so, we will demonstrate that the Hawking spectrum must be Planckian due to the presence of a Stokes line, without using the unitarity constraint. 

There are alternative derivations of the Planck spectrum from on-shell amplitudes, see, e.g., \cite{Aoki:2025ihc, Akpinar:2026oni}. These papers compute the total emission probability using three-particle massive amplitudes. By requiring that the macroscopic average of the massive spin states is zero, they show the distribution must be thermal. The temperature for the black hole is then derived from the gray-body factor by requiring the horizon saturates a unitarity bound, that the total absorption probability never exceeds one. In our paper, we propose a method for deriving the temperature directly from the calculation of the Bogoliubov coefficient.

The Stokes phenomenon we discuss in this paper bears striking resemblance to that discussed in the case of acoustic sound waves in \cite{Farrell:2022mzv}. In our paper, we apply their use of exponential coordinates to black holes instead, but in \autoref{s:AcousticRadiation} we strengthen the connection by illustrating how the subtleties having to do with the eikonal approximation can also be seen in the case of acoustic sound waves. The study of Stokes phenomenon for black holes specifically is also not new. This was explored for example in \cite{Dumlu:2020wvd, Enomoto:2022mti}. In those cases, however, they work in position space. We instead work in energy space, which is better suited for an amplitudes calculation.

\textbf{Outline of the Paper.} In \autoref{s:AcousticRadiation}, we explore a simplified toy model which illustrates the difference between employing a weak-field approximation followed by a geometric optics and the reverse. Then, in \autoref{s:HawkingSpectrum}, we compute the Bogoliubov coefficient for a black hole that is spherically symmetric whose geometry admits a Kerr--Schild form. The method is based on the setup in \cite{Aoude:2024sve}, but uses geometric optics approximations such as those used in path integral approaches \cite{Parikh:1999mf, Harlow:2014yka, Padmanabhan:2016xjk}. Finally, in \autoref{s:StokesLine}, we compute the Bogoliubov coefficients in the saddle point approximation and demonstrate that, using complex coordinates for the integral, the Stokes line in the complex energy plane is related to the temperature an observer measures in that black hole background. We discuss these results as well as next steps in \autoref{s:Discussion}.

\textbf{Notations and Conventions.} Throughout these notes, we will use the QFT conventions of \cite{Srednicki:2007}, with units $c = \hbar = 1$. In particular, we will use the mostly plus metric signature $\left(-, +, \dots, +\right)$ in $d$ spacetime dimensions. Our Fourier transform conventions are
\begin{equation}
\tilde{f}\left(k\right) = \int f\left(x\right) \e^{-ik \cdot x} \dd[d]{x} \qquad \text{and} \qquad f\left(x\right) = \int \tilde{f}\left(k\right) \e^{ik \cdot x} \frac{\dd[d]{k}}{\left(2\pi\right)^d}\,.
\end{equation}
We will also define the on-shell phase space measure to be
\begin{equation}
\label{eq:PhaseSpaceDef}
\dd{\Phi}\left(p\right) = \frac{\dd[d]{p}}{\left(2\pi\right)^d} 2\pi\delta\left(p^2\right)\Theta\left(p^0\right). 
\end{equation}
Finally, we will use $\alpha$ and $\beta$ to define the Bogoliubov coefficients, as discussed in the context of QFT in curved spacetime, e.g. in \cite{Parker:2009}. We will use $\mathcal{A}$ and $\mathcal{B}$ to denote the generalized amplitudes of \cite{Aoude:2024sve}, which are related to the Bogoliubov coefficients by a constant prefactor. 

\section{Toy Model: Acoustic Black Hole}
\label{s:AcousticRadiation}

In this section, we study Hawking radiation in a more familiar
%\DM{more familiar?} 
setting that will help elucidate the interplay between the eikonal and weak field approximations. A toy model that mirrors many of the effects seen in black holes is that of fluid flow. Physically, the fluid velocity changes with the cross-sectional area of the fluid container. If the velocity exceeds the local speed of sound, there will be an acoustic horizon for sound waves. For a review see \cite{Visser:1997ux} as well as \cite{Barcelo:2005fc, Barcelo:2026vlr} for a clear depiction of the acoustic horizon in terms of upstream flow. This was used to study the possibility of observing a Hawking spectrum of phonons in fluid systems \cite{Dave:2022wgk} as well as to elucidate features of tidal Love numbers \cite{DeLuca:2024nih, DeLuca:2025zqr}. 

For computational simplicity, we will work in coordinates where time, $t$, is null. It is related to the time used in e.g. \cite{Visser:1997ux, Barcelo:2005fc, Barcelo:2026vlr}, $t_0$, by a coordinate transformation $t = t_0 + \dd{x}/\left(c - v\right)$. In these coordinates, the metric is
\begin{equation}
\dd{s}^2 = -\left(c^2 - v^2\left(x\right)\right)\dd{t}^2 + 2 c \dd{t} \dd{x},
\end{equation} 
where $v\left(x\right)$ is the fluid velocity and the sound speed is $c$. The action for a sound mode, $\phi$, in these coordinates is
\begin{equation}
\label{eq:AcousticLagrangian}
S = -\frac{1}{2}\int \left[\frac{2}{c}\partial_t \phi\, \partial_x \phi + \left(\partial_x \phi\right)^2 - \frac{v^2}{c^2} \left(\partial_x \phi\right)^2\right] \dd[2]{x}. 
\end{equation}
The equation of motion from this action is
\begin{equation}
\left[\frac{2}{c}\partial_t \partial_x + \partial_x^2 - \partial_x\left(\frac{v^2}{c^2} \partial_x\right)\right]\phi = 0.
\end{equation}
In order to solve this equation, we assume the form 
\begin{equation}
\phi = \operatorname{e}^{-i E t + is\left(x\right)},
\end{equation}
and work in the geometric optics approximation $E \gg \partial_x v$. With that said, the solution in the geometric optics approximation in this case is exact. This can be seen by noting that under a further coordinate transformation of $x$, the metric is conformally flat. In 2d, the solutions to a free field in a conformally flat background are plane waves. In this case, the real part of the equation of motion for $s\left(x\right)$ is
\begin{equation}
s^\prime\left(x\right)\left(s^\prime\left(x\right) - \frac{2c E}{c^2 - v^2}\right) = 0.
\end{equation}
Therefore, the two possible solutions are constant $s\left(x\right)$ and
\begin{equation}
\label{eq:NonPert Eikonal}
s\left(x\right) = \int^x \frac{2c E}{c^2 - v^2\left(x^\prime\right)} \dd{x^\prime}.
\end{equation}
The imaginary part of the equations of motion are then satisfied exactly. Physically, the second solution corresponds to the upstream solution shown in fig. 6 of \cite{Barcelo:2005fc, Barcelo:2026vlr}. The acoustic horizon appears for $v\left(x\right) \sim c$ and appears as a pole in $s^\prime\left(x\right)$. We select out this solution as opposed to the constant solution by requiring that the solution is a plane wave as $x \rightarrow -\infty$. 

One could then wonder whether this statement could be recovered from a perturbative amplitudes calculation. For the action in \autoref{eq:AcousticLagrangian}, the diagrammatic expansion is an expansion in the weak field limit, $v/c \ll 1$. For simplicity, and to ensure there are well-defined asymptotic states, we take $v\left(x\right) = 0$ as $x \rightarrow \pm \infty$. The Feynman rule for the leading interaction diagram is
\begin{subequations}
\begin{align}
    \begin{gathered}
    \vspace{-0.5cm}
    \begin{tikzpicture}
        \draw[line width=0.3mm] (-1.5, 0) -- (-0.15, 0);
        \draw[line width=0.3mm] (0.15, 0) -- (1.5, 0);
        \draw[line width=0.3mm] (0, 0) node[cross=0.15cm, BrickRed] {};
        \draw[line width=0.3mm, BrickRed] (0, 0) circle (0.15cm);
        \draw[line width=0.3mm, -Latex] (-1.35, -0.2) -- node[anchor=north] {$\left(E, k\right)$} (-0.3, -0.2);
        \draw[line width=0.3mm, -Latex] (0.3, -0.2) -- node[anchor=north] {$\left(E^\prime, k^\prime\right)$} (1.35, -0.2);
        \draw[line width=0.3mm, -Latex, BrickRed] (0, 0.75) -- node[anchor=east] {$q$} (0, 0.2);
    \end{tikzpicture}
    \end{gathered} \: \: &= \frac{ik k^\prime}{c^2} 2\pi\delta\left(E^\prime - E\right) \int \dd{x} \operatorname{e}^{-i q x} v^2\left(x\right),\\ 
    &= \frac{i k k^\prime}{c^2} \left(\tilde{v} \ast \tilde{v}\right)_q 2\pi \delta\left(E^\prime - E\right),
\end{align}
\end{subequations}
where $q = k^\prime - k$ and $\ast$ denotes a convolution. Recall that when the coupling is a fixed background field instead of a constant, the momentum conserving $\delta$-distribution is replaced with a Fourier transform. This can be thought of as encoding the fact that the external field is injecting momentum into the particle. Then, the propagator is 
\begin{equation}
    \begin{gathered}
    \vspace{-0.75cm}
    \begin{tikzpicture}
        \draw[line width=0.3mm] (-1.6, 0) -- (0.1, 0);
        \draw[line width=0.3mm, -Latex] (-1.35, -0.2) -- node[anchor=north] {$\left(E, k\right)$} (-0.15, -0.2);
    \end{tikzpicture}
    \end{gathered} \: \: = \frac{-i}{k\left(k - 2E/c\right)}.
\end{equation}
Notice that because the $t$ coordinate is technically the advanced time, the denominator is not the standard $E^2 - k^2$. 

With this set up, we would like to compute $\phi\left(x\right)$ directly from a scattering amplitudes calculation. Our boundary condition is that as $x \rightarrow -\infty$
\begin{equation}
\phi\left(x\rightarrow -\infty\right) = \operatorname{e}^{-i E \left(t - 2 x/c\right)},
\end{equation}
since $v\left(x\right) \rightarrow 0$ there. The $n^{\text{th}}$-order contribution to the one-point function is then given by the Lippmann--Schwinger iterative series, i.e. in eq. (4.34) of \cite{Aoude:2024sve}:
\begin{equation}
\phi^{\left(n\right)}\left(x, k_0\right) = \int \frac{\dd{k_n}}{2\pi} \dots \frac{\dd{k_1}}{2\pi} \operatorname{e}^{-i E t + i k_n x} \frac{k_n \left(\tilde{v}\ast \tilde{v}\right)_{k_n - k_{n - 1}} k_{n - 1}}{c^2 k_n\left(k_n - k_0\right)} \dots \frac{k_1 \left(\tilde{v} \ast \tilde{v}\right)_{k_1 - k_0} k_0}{c^2 k_1\left(k_1 - k_0\right)}.
\end{equation}
Here, we have already used energy conservation at every vertex. Note that $k_0 = 2E/c$, the initial momentum. Diagrammatically, this is given by
\begin{equation}
   \phi^{\left(n\right)}\left(x, k_0\right) = \: \: \begin{gathered}
    \vspace{-0.5cm}
    \begin{tikzpicture}
        \draw[line width=0.3mm] (-1.5, 0) -- (-0.15, 0);
        \draw[line width=0.3mm] (0.15, 0) -- (1.5, 0);
        \draw[line width=0.3mm] (0, 0) node[cross=0.15cm, BrickRed] {};
        \draw[line width=0.3mm, -Latex, BrickRed] (0, 0.75) -- node[anchor=east] {} (0, 0.2);
        \draw[line width=0.3mm, BrickRed] (0, 0) circle (0.15cm);
        \draw[line width=0.3mm, -Latex] (-1.35, -0.2) -- node[anchor=north] {$k_0$} (-0.3, -0.2);
        \draw[line width=0.3mm, -Latex] (0.3, -0.2) -- node[anchor=north] {$k_1$} (1.35, -0.2);
        \draw[line width=0.3mm, -Latex] (2.15, -0.2) -- node[anchor=north] {$k_{n - 1}$} (3.2, -0.2);
        \draw[line width=0.3mm, -Latex] (3.8, -0.2) -- node[anchor=north] {$k_n$} (4.85, -0.2);
        \node at (1.775, 0) {$\dots$};
        \draw[line width=0.3mm] (2, 0) -- (3.35, 0);
        \draw[line width=0.3mm] (3.65, 0) -- (5, 0);
        \draw[line width=0.3mm] (3.5, 0) node[cross=0.15cm, BrickRed] {};
        \draw[line width=0.3mm, BrickRed] (3.5, 0) circle (0.15cm);
        \draw[line width=0.3mm, -Latex, BrickRed] (3.5, 0.75) -- node[anchor=east] {} (3.5, 0.2);
        \filldraw (5, 0) circle (0.075cm);
        \node at (5, 0.25) {$x$};
        \filldraw[OliveGreen] (-1.425, 0.075) rectangle (-1.575, -0.075); 
    \end{tikzpicture}
    \end{gathered} \: \: .
\end{equation}
The left square vertex indicates the source giving the original momentum $k_0$. The right black dot indicates that one should Fourier transform back to position space. 

Instead of resumming the entire series, we will first evaluate the lower order single-vertex contribution. We will then argue that the entire series can be resummed via eikonal exponentiation. The lowest order contribution is
\begin{equation}
\phi^{\left(1\right)}\left(x, k_0\right) = \frac{k_0 \operatorname{e}^{-i E t}}{c^2} \int \frac{\operatorname{e}^{i k x} \operatorname{e}^{-i q x^\prime}}{\left(k - k_0\right) + i\varepsilon} v^2\left(x^\prime\right) \frac{\dd{k}}{2\pi} \dd{x^\prime}.
\end{equation}
Here, we have restored the $i\varepsilon$ to enforce the proper boundary conditions. Transforming the integration variable from $k$ to $q = k - k_0$ and using 
\begin{equation}
\int \frac{\operatorname{e}^{-i q x}}{q + i\varepsilon} \frac{\dd{q}}{2\pi} = -i\Theta\left(x\right),
\end{equation}
where $\Theta$ is the Heaviside distribution, we find 
\begin{equation}
\phi^{\left(1\right)}\left(x, k_0\right) = \operatorname{e}^{-i E \left(t - 2x/c\right)} \frac{-2 i E}{c^3} \int^{x} v^2\left(x^\prime\right) \dd{x^\prime} .
\end{equation}
Note, here, that $k_j - k_0 = q_1 + \dots + q_j$ exactly. We use the eikonal approximation in the numerator, $k_{j - 1} \approx k_0$. Then, each integral over $q$ gives successive $\Theta$ distributions. Therefore, using 
\begin{equation}
\int \dd[n]{x} \Theta\left(x_1 < \dots < x_n < x\right) \prod_{j} \left(-\frac{i k_0}{c^2} v^2\left(x_j\right)\right) = \frac{1}{n!}\left(-\frac{i k_0}{c^2} \int v^2\left(x^\prime\right) \dd{x^\prime}\right)^n,
\end{equation}
we find that the tree-level result exponentiates
\begin{equation}
\phi\left(x, k_0\right) = \operatorname{e}^{-i E t + i s_{\text{pert}}\left(x\right)},
\end{equation}
where
\begin{equation}
s_{\text{pert}}\left(x\right) = \frac{2E}{c}x + \frac{2E}{c^3} \int^x v^2\left(x^\prime\right) \dd{x^\prime}. 
\end{equation}
This only agrees with \autoref{eq:NonPert Eikonal} in the weak field expansion $v^2/c^2$. The only difference between these approaches is the use of the weak-field limit when isolating the leading eikonal. Once we resum all the diagrams, we supposedly leave the weak-field limit, but this is still not able to recover the full result.

To make the physics concrete, we choose a specific profile for the velocity field, 
\begin{equation}
\label{eq:Tangherlini Profile}
v^2\left(x\right) = c^2\left(\frac{x_0}{x}\right)^n,
\end{equation}
where $n$ is a positive integer and $x_0$ plays the role of the location of the horizon. The solution for $s\left(x\right)$ in \autoref{eq:NonPert Eikonal} develops a non-trivial imaginary part near $x_0$,
\begin{equation}
\label{eq:AcousticT}
\frac{1}{c^2}\oint_{|x - x_0| = \epsilon} \frac{2 E c}{1 - \left(x_0/x\right)^n} \frac{\dd{x}}{2\pi i} = \frac{2 E x_0}{c n}. 
\end{equation}
As we can see, if we evaluated the residue near $x_0$, we would get zero. But, there is a new singularity located at $x = 0$. There, 
\begin{equation}
\frac{2E}{c} \oint_{|x| = \epsilon}  \left(\frac{x_0}{x}\right)^n \frac{\dd{x}}{2\pi i} = \frac{2E x_0}{c} \delta_{n, 1}.
\end{equation}
This evaluation of the residue agrees with \autoref{eq:AcousticT} for $n = 1$, but has no residue for $n \geq 2$. The extraction of the non-perturbative information, contained in this residue, is therefore in general not possible to obtain at any finite order in perturbation theory, 
\begin{equation}
\oint s^\prime\left(x\right) \frac{\dd{x}}{2\pi i} \neq \oint s_{\text{pert}}^\prime\left(x\right) \frac{\dd{x}}{2\pi i}.
\end{equation}
%\DM{Should this be the derivative of s in the integral?} 
In the next section, we will apply this understanding to the case of higher dimensional black holes and we will see that this profile is chosen to mimic the features seen in Hawking radiation for a higher dimensional Schwarzschild black hole, called Schwarzschild--Tangherlini. We will see that the $n = 1$ case where there is indeed an equivalent residue corresponds to a 4d Schwarzschild black hole, but that this equivalence does not carry over to the higher dimensional generalization nor to other black holes, such as charged black holes. 

\section{The Kerr--Schild Hawking Amplitude}
\label{s:HawkingSpectrum}

In the previous section, we demonstrated that the weak-field and eikonal limits do not commute in general. In this section, we derive the Hawking spectrum via the Hawking amplitude defined by \cite{Aoude:2024sve}, but ensuring to only employ a geometric optics approximation. We then demonstrate that when taking the weak field limit first and then the eikonal limit, the resulting Hawking amplitude still retains the weak field approximation even after resummation. For simplicity, we restrict our focus to black holes whose metric is spherically symmetric and can be written in Kerr--Schild form,
\begin{equation}
\label{eq:Kerr-Schild Form}
g_{\mu\nu} = \eta_{\mu\nu} + h\left(r\right) \Theta\left(t + r\right) \ell_\mu \ell_\nu,
\end{equation}
where the flat metric, $\eta$, is most usefully expressed in spherical coordinates and $\ell^\mu$ is null with respect to both $\eta$ and $g$ and given by
\begin{equation}
\ell_\mu \ell_\nu \dd{x}^\mu \dd{x}^\nu = \left(\dd{t} + \dd{r}\right)^2.
\end{equation}
We include a Heaviside $\Theta$-function in \autoref{eq:Kerr-Schild Form} to indicate we are working in a \textit{collapse background} where the black hole forms instantaneously. Before its formation, the spacetime is flat and afterwards, it is given by the Kerr--Schild metric with potential $h\left(r\right)$. The example Kerr--Schild potentials we use in this section are
\begin{subequations}
\begin{align}
\label{eq:Ex1}
\text{Schwarzschild--Tangherlini: } h\left(r\right) &= \left(\frac{r_s}{r}\right)^{d - 3},\\
\text{Reissner--Nordstr\"{o}m: } h\left(r\right) &= \left(\frac{r_s}{r}\right)^{d - 3} - \left(\frac{r_q}{r}\right)^{2\left(d - 3\right)}.
\label{eq:Ex2}
\end{align}
\end{subequations}
where Schwarzschild--Tangherlini describes a spherically symmetric $d$-dimensional neutral and static black hole and Reissner--Nordstr\"{o}m describes the same black hole endowed with charge $r_q^{d-3}/\sqrt{G}$.

Two useful technical properties of Kerr--Schild coordinates is that the inverse metric is exactly
\begin{equation}
g^{\mu\nu} = \eta^{\mu\nu} - h\left(r\right)\Theta\left(t + r\right)\ell^\mu \ell^\nu
\end{equation}
and the metric determinant in Cartesian coordinates is $\det g_{\mu\nu} = -1$. Normally, the inverse metric gets contributions at \textit{all} orders in the function $h$, but here it stops at linear order. Hence, the action for a scalar field in this background decomposes, without approximation, into a free piece and one interaction piece, 
\begin{equation}
\label{eq:FullAction}
S = S_{\text{free}} + \frac{1}{2} \int h\left(r\right) \Theta\left(t + r\right)\left(\partial_t \phi - \partial_r \phi\right)^2 \dd[d]{x}.
\end{equation}
This is the action we will use for the remainder of the section. 

\subsection{The Hawking Spectrum from Amplitudes}
As proposed by \cite{Aoude:2024sve, Aoude:2026bek}, the Bogoliubov coefficients can be expressed, up to an overall factor, in terms of an amplitudes calculation using the KMOC formalism. For a review of the KMOC formalism with application to classical gravity computations, see \cite{Kosower:2022yvp}. To begin, they define the \textit{Hawking Amplitude}\footnote{There is a difference in naming convention between \cite{Aoude:2024sve} and this paper. To clarify, in this paper we refer to the Hawking amplitude by the given $S$-matrix element and the Bogoliubov coefficient by the overlap of wave functions. They differ by an overall constant factor.},
\begin{equation}
\mathcal{A} \equiv \mel{p^\prime}{S}{\psi},
\end{equation}
where $\ket{\psi}$ is spherically symmetric on-shell state of infalling radiation,
\begin{equation}
\ket{\psi} \equiv \int \dd{\Phi}\left(p\right) \ket{p} \int \dd{v} \e^{iEv} \varphi\left(v\right). 
\end{equation}
Here, $\varphi\left(v\right)$ is a wavepacket describing the in-falling radiation. 

By directly plugging in the state $\ket{\psi}$, we find
\begin{equation}
\label{eq:Generalized Amplitude to S Matrix}
\mathcal{A} = \int \dd{v} \varphi\left(v\right)\int \dd{\Phi}\left(p\right) \e^{ip \cdot b}\mel{p^\prime}{S}{p},
\end{equation}
where $b^\mu = \left(v, \vb{0}\right)$. Therefore, the computation of the Hawking amplitude reduces to the computation of a one-to-one transition amplitude. The operator whose matrix elements give this amplitude will be called $D$. Diagrammatically, this is given by
\begin{subequations}
\begin{align}
\label{eq:Resummation}
D &= \: \: \sum_{\text{insertions}}
\begin{gathered}
    \vspace{-0.5cm}
    \begin{tikzpicture}
        \draw[line width=0.3mm] (-1.5, 0) -- (-0.15, 0);
        \draw[line width=0.3mm] (0.15, 0) -- (1.5, 0);
        \draw[line width=0.3mm] (0, 0) node[cross=0.15cm, RoyalBlue] {};
        \draw[line width=0.3mm, RoyalBlue] (0, 0) circle (0.15cm);
        \draw[line width=0.3mm, -Latex] (-1.35, -0.2) -- node[anchor=north] {$p$} (-0.3, -0.2);
        \draw[line width=0.3mm, -Latex] (3.8, -0.2) -- node[anchor=north] {$p^\prime$} (4.85, -0.2);
        \node at (1.775, 0) {$\dots$};
        \draw[line width=0.3mm] (2, 0) -- (3.35, 0);
        \draw[line width=0.3mm] (3.65, 0) -- (5, 0);
        \draw[line width=0.3mm] (3.5, 0) node[cross=0.15cm, RoyalBlue] {};
        \draw[line width=0.3mm, RoyalBlue] (3.5, 0) circle (0.15cm);
    \end{tikzpicture}
    \end{gathered} \: \: ,\\[0.2cm]
&= D_0 + D_0 V D_0 + \dots + \left(D_0 V D_0\right)^n + \dots,\\
&= \frac{1}{D_0^{-1} - V},
\end{align}
\end{subequations}
where $D_0$ is the free propagator operator and $V$ is the interaction operator associated to the second term in \autoref{eq:FullAction}. Since the free action is Minkowski, the Hilbert space states coincide with those of flat space field theory. Evaluated in the position basis, this operator is the two-point function. In other words, by the LSZ reduction formula,  
\begin{equation}
\mel{p^\prime}{S}{p} = \int \dd[d]{x} \dd[d]{x^\prime} \e^{ip^\prime x^\prime} \e^{-ip x} \mel{x^\prime}{-iD}{x},
\end{equation}
where the on-shell limit is understood. Since $D$ is an inverse operator, it can be expressed in Schwinger parameter form. The matrix element of $D^{-1}$ is then equivalent to a quantum mechanical transition probability, with $D^{-1}$ playing the role of the Hamiltonian. This can be expressed as a path integral:   
\begin{subequations}
\begin{align}
\mel{x^\prime}{-iD}{x} &=  \mel{x^\prime}{\int_{0}^{\infty} \dd{T} \e^{-i D^{-1} T}}{x}, \\
&= \int_0^\infty \dd{T} \int \mathcal{D} x \mathcal{D} P \, \e^{i \int_{0}^{T} \left(P_\mu \dot{x}^\mu - D^{-1}\right)\dd{\tau} }.
\end{align}
\end{subequations}
This is commonly referred to as the \textit{worldline action}. 

Up to this point, we have computed everything exactly. We would like to compute the Hawking amplitude in the geometric optics approximation, $\abs{q} \ll E$. To do so, we use a saddle point approximation on the time and path integrals. This enforces $D^{-1} = 0$ as well as the selection of the classical path, meaning the one-to-one transition amplitude reduces to 
\begin{equation}
\mel{p^\prime}{S}{p} = \int \dd[d]{x} \dd[d]{x^\prime} \e^{ip^\prime x^\prime} \e^{-ipx} \e^{i\int_{x}^{x^\prime} P \cdot \dd{x}}.
\end{equation}
Plugging this and the definition of the phase space measure given by \autoref{eq:PhaseSpaceDef} into the Hawking amplitude, we find
\begin{equation}
\mathcal{A} = \int \dd{v} \varphi\left(v\right) \int \frac{\dd[d]{p}}{\left(2\pi\right)^d} \dd[d]{x} \dd[d]{x^\prime} 2\pi \delta (p^2) \e^{-ip\left(x - b\right)} \e^{ip^\prime x^\prime} \e^{i\int_{x}^{x^\prime} P \cdot \dd{x}}.
\end{equation}
We can additionally promote the $\delta$-distribution to be an exponential with $\lambda$ parameterizing the exponential, 
\begin{equation}
\mathcal{A} = \int \dd{v} \varphi\left(v\right) \int \frac{\dd[d]{p}}{\left(2\pi\right)^d} \dd{\lambda} \dd[d]{x} \dd[d]{x^\prime} \e^{i\lambda p^2} \e^{i p^\prime x^\prime} \e^{-ip \left(x - b\right)} \e^{i\int_{x}^{x^\prime} P \cdot \dd{x}}. 
\end{equation}
We can employ a saddle point approximation on the remaining integrals. The saddle point on $x$ and $x^\prime$ respectively define $P\left(x\right) = p$ and $P\left(x^\prime\right) = p^\prime$. The saddle point in $\lambda$ returns the constraint $p^2 = 0$. Finally, the saddle point in $p$ yields $x = b + 2\lambda p$. Since $b^\mu = \left(v, 0\right)$, this implies
\begin{equation}
t = v + 2 E \lambda \quad \text{and} \quad r = \abs{x^i} = 2 E \abs{\lambda}. 
\end{equation}
By the Heaviside function in \autoref{eq:FullAction}, we find that for positive $\lambda$, $t + r = v + 4E\lambda > 0$. This then implies that $r > -v/2$. These bounds coincide with \cite[eq. (3.19)]{Aoude:2024sve}. As a result the Hawking amplitude is,
\begin{equation}
\mathcal{A} = \int \dd{v} \varphi\left(v\right) \e^{i\int_{r = -v/2}^{r = \infty} \left(p^\prime - P\right) \cdot \dd{x}} ,
\end{equation}
with the lower bound on the radial coordinate being $-v/2$ while the upper bound is infinity. Finally, we impose the on-shell condition for $P$, $g^{\mu\nu} P_\mu P_\nu = 0$, which states
\begin{equation}
g^{\mu\nu} P_\mu P_\nu = 0 \Rightarrow P_r\left[2E + \left(1 - h\right) P_r\right] = 0,
\end{equation}
where $P_v = E$ as fixed by the saddle point condition $P\left(x^\prime\right) = p^\prime$. There are two solutions to this equation, $P_r = 0$ and 
\begin{equation}
P_r = -\frac{2E}{1 - h}.
\end{equation}
By the condition from the outgoing saddle, we must have $P_r \neq 0$. As such, we plug the other solution into the exponential. Since $\left(p^\prime - P\right)_v = 0$, the Hawking amplitude is simply
\begin{equation}
\label{eq:Full Hawking A}
\mathcal{A} = \int \dd{v} \varphi\left(v\right) \e^{i E v} \exp\left[i E \int_{-v/2}^{\infty} \frac{2h\left(r^\prime\right)}{1 - h\left(r^\prime\right)} \dd{r^\prime} \right], 
\end{equation}
where the linear in $v$ term comes from the lower bound of the $x$ integral. 

At this point, we specify the form of the infalling radiation, as in \cite{Aoude:2024sve} to be 
\begin{equation}
\varphi\left(v\right) = \frac{2\pi}{E_0} \e^{-i E_0 v}.
\end{equation}
With this, the Hawking amplitude takes the form
\begin{equation}
\label{eq:Bogoliubov from Amplitude}
\mathcal{A} = \frac{2\pi}{E_0} \int \dd{v} \e^{-i E_0 v} \e^{i E u\left(v\right)},
\end{equation}
with $u\left(v\right)$ defined as
\begin{equation}
\label{eq:Ray Tracing from Amplitudes}
u\left(v\right) = v + \int_{-v/2}^{\infty} \frac{2h\left(r^\prime\right)}{1 - h\left(r^\prime\right)} \dd{r^\prime}.
\end{equation}
Notice that the location of the horizon appears now as the location of the singularity in the Hawking amplitude, $1 - h = 0$. This exponent, derived from the one-to-one transition amplitude, is equivalent to the ray tracing relationship obtained in \autoref{eq:Ray Tracing Final}. This should not be surprising since the ability to use rays is equivalent to the geometric optics approximation we used here.

To close out this subsection, we state the ray tracing results for \autoref{eq:Ex1} and \autoref{eq:Ex2} respectively:
\begin{subequations}
\begin{align}
\label{eq:ST u}
u_{\text{ST}}\left(v\right) &= v - \frac{2 r_s}{d - 3} \sum_{k = 0}^{d - 4} \e^{\frac{2\pi i k}{d - 3}} \log\left(1 + \frac{2 r_s}{v} \e^{\frac{2\pi i k}{d - 3}}\right),\\
u_{\text{RN}}\left(v\right) &= v - \frac{2}{\left(d - 3\right)\left(r_+^{d - 3} - r_-^{d - 3}\right)}\Big[r_+^{d - 2} \sum_{k = 0}^{d - 4} \e^{\frac{2\pi i k}{d - 3}} \log\left(1 + \frac{2r_+}{v}\e^{\frac{2\pi i k}{d - 3}}\right) \label{eq:RN u} \\ 
&\hspace{4cm} - r_-^{d - 2}\sum_{k = 0}^{d - 4} \e^{\frac{2\pi i k}{d - 3}} \log\left(1 + \frac{2 r_-}{v} \e^{\frac{2\pi i k}{d - 3}}\right)\Big]. \nonumber 
\end{align}
\end{subequations}
In the second case, $r_+$ and $r_-$ correspond to the locations of the outer and inner horizons respectively,
\begin{equation}
r_{\pm}^{d - 3} = \frac{1}{2}\left[r_s^{d - 3} \pm \sqrt{r_s^{2\left(d - 3\right)} - 4 r_q^{2\left(d - 3\right)}}\,\right].
\end{equation}
In four spacetime dimensions, the integrals are generically divergent. The divergences are referred to as Coulomb phases and can be regularized with a length scale $\ell > 0$. So, for $d = 4$, 
\begin{subequations}
\begin{align}
\label{eq:4dSchw Result}
u_{\text{S}}\left(v\right) &= v - 2r_s \log\left(\frac{v + 2 r_s}{-\ell}\right),\\
u_{\text{RN}}\left(v\right) &= v - \frac{2}{r_+ - r_-}\left[r_+^2 \log\left(\frac{v + 2 r_+}{-\ell}\right) - r_-^2 \log\left(\frac{v + 2 r_-}{-\ell}\right)\right]. 
\label{eq:4dRN}
\end{align}
\end{subequations}

\subsection{The Case of Reversed Limits}
\label{s:The Case of Reversed Limits}
In the previous subsection, we obtained a result for the leading eikonal contribution to the Hawking amplitude. In \autoref{eq:Resummation}, we summed all insertions of the interaction vertex and then used an eikonal approximation on the resulting one-to-one transition amplitude. Here, we reverse the order of operations. We instead use the eikonal approximation \textit{before} resumming all insertions. While the answer still exponentiates, as pointed out by \cite{Aoude:2024sve}, the analytic structure of the result is different. Physically, this implies a different temperature measured by an observer far away. For concreteness, we will work with a 4d collapsing Reissner--Nordstr\"{o}m black hole.

To begin, let's return to \autoref{eq:Generalized Amplitude to S Matrix}, which says the Hawking amplitude can be computed by integrating over a one-to-one transition amplitude. In turn, the one-to-one transition amplitude is given by the sum of insertions of the Kerr--Schild potential, $h\left(r\right)$, just as in \autoref{eq:Resummation}, but now evaluated in momentum space:
\begin{equation}
\mel{p^\prime}{S}{p} = \: \: \sum_{\text{insertions}}
\begin{gathered}
    \vspace{-0.5cm}
    \begin{tikzpicture}
        \draw[line width=0.3mm] (-1.5, 0) -- (-0.15, 0);
        \draw[line width=0.3mm] (0.15, 0) -- (1.5, 0);
        \draw[line width=0.3mm] (0, 0) node[cross=0.15cm, RoyalBlue] {};
        \draw[line width=0.3mm, RoyalBlue] (0, 0) circle (0.15cm);
        \draw[line width=0.3mm, -Latex] (-1.35, -0.2) -- node[anchor=north] {$p$} (-0.3, -0.2);
        \draw[line width=0.3mm, -Latex] (3.8, -0.2) -- node[anchor=north] {$p^\prime$} (4.85, -0.2);
        \node at (1.775, 0) {$\dots$};
        \draw[line width=0.3mm] (2, 0) -- (3.35, 0);
        \draw[line width=0.3mm] (3.65, 0) -- (5, 0);
        \draw[line width=0.3mm] (3.5, 0) node[cross=0.15cm, RoyalBlue] {};
        \draw[line width=0.3mm, RoyalBlue] (3.5, 0) circle (0.15cm);
    \end{tikzpicture}
    \end{gathered} \: \: . 
\end{equation}
Each insertion carries a power of Newton's constant $G$, since both $r_s$ and $r_q^2$ are proportional to $G$. Each diagram will then have its own eikonal expansion parameterized by $\eta$. We therefore write the one-to-one transition amplitude as
\begin{equation}
\label{eq:Aexpansion}
\mel{p^\prime}{S}{p} = \sum_{L = 0}^{\infty} \sum_{k = -L}^{\infty} G^{L + 1} \eta^{k - 1} i A_L^{\left(k\right)}. 
\end{equation}

The diagram corresponding to a single insertion of $h$ is determined by the action in \autoref{eq:FullAction} to be
\begin{equation}
A_0\left(q\right) = \int \dd[4]{x} \e^{iq\cdot x} p_\mu h^{\mu\nu}\left(x\right) \left(p - q\right)_\nu,
\end{equation}
where for simplicity, we define $h_{\mu\nu} = g_{\mu\nu} - \eta_{\mu\nu}$. The appearance of an integral over space is expected -- the fields couple to an external background $h_{\mu\nu}$, hence the usual momentum conserving $\delta$-distribution is replaced by the Fourier transform of the background field, which injects momentum $q$ into the particle. At this order, the Hawking amplitude, $\mathcal{A}$, is expressible in terms of the \textit{impact-parameter} transform of $A_0$:
\begin{equation}
\mathcal{A} = \int \dd{v} \varphi\left(v\right) \e^{ip^\prime \cdot b\left(v\right)} \tilde{A}_0\left(b\right),
\end{equation}
where, following the same steps of exponentiating the $\delta$-function as in the previous subsection, 
\begin{equation}
\tilde{A}_0\left(b\right) = \int \frac{\dd[4]{q}}{\left(2\pi\right)^4} \dd[4]{x} \dd{\lambda} \e^{i\left(x - b - 2 \lambda p\right) \cdot q + i\lambda q^2} p_\mu h^{\mu\nu}\left(x\right) \left(p - q\right)_\nu. 
\end{equation}
By dropping the $q^2$ term in the exponent, we will be evaluating the leading eikonal contribution at this order. Evaluating the integral, we find
\begin{equation}
\tilde{A}_0^{\left(0\right)}\left(b\right) = - 2 E r_s \log\left(\frac{v}{-\ell}\right) + 2 E r_s \log\left(\frac{4 E \lambda_{\infty}}{\ell}\right) + \frac{4 E r_q^2}{v}, 
\end{equation}
where $\lambda_{\infty}$ regulates the divergence from the upper bound of the $\lambda$ integral. This is the Coulomb phase and, as in the previous subsection, we will drop it in remaining calculations.

This procedure can be carried over to an arbitrary number of insertions of the field $h_{\mu\nu}$. In that case, as shown in \cite{Aoude:2024sve}, the leading eikonal contribution at a given order in number of insertions of $h_{\mu\nu}$ is a \textit{power} of the lowest order result, 
\begin{equation}
i\tilde{A}_{L}^{\left(-L\right)} = \frac{1}{\left(L + 1\right)!} \left(i\tilde{A}_0^{\left(0\right)}\right)^L. 
\end{equation}
Given this, the sum over all insertions is then an exponential of the leading result, giving the following expression for the Hawking amplitude, 
\begin{equation}
\label{eq:leadingEikExp}
\mathcal{A} = \int \dd{v} \varphi\left(v\right) \e^{ip\cdot b\left(v\right)} \exp\left[-2i E r_s \log\left(\frac{v}{-\ell}\right) + \frac{4 i E r_q^2}{v}\right]. 
\end{equation}
We emphasize the difference between this result and the one in \autoref{eq:4dRN}. In $v$, there is indeed a log with a branch cut, albeit at a different location. However, the coefficient of this log is different and hence, at this order would predict different physics. As we will show in the next section, this log coefficient contains information about the temperature of the black hole. 

\subsection{Matching to the Eikonal Exponent}
\label{s:Matching to the Eikonal Exponent}
What we have demonstrated in the previous subsection is that the temperature can be recovered from exponentiating the leading vertex only for a 4d Schwarzschild black hole. For more general black holes, including both Schwarzschild--Tangherlini and Reissner--Nordstr\"{o}m, one needs the subleading eikonal contributions to $\tilde{A}\left(b\right)$.

As reviewed in \cite{Aoude:2024sve}, however, there are different notions of the eikonal. In particular, the eikonal exponent can be related to the amplitude in impact parameter space by the common expression for eikonal exponentiation \cite{Damgaard:2021ipf,DiVecchia:2023}:
\begin{equation}
\label{eq:eikExp}
    \e^{i\chi/\eta}\left(1 + i \Delta\right) = 1+i\tilde{A}.
\end{equation}
The eikonal exponent, $\chi$, and the remainder, $\Delta$, additionally admit weak field expansions 
\begin{equation}
\chi = \sum_{n = 0}^{\infty} G^{n + 1} \chi_n \quad \text{and} \quad \Delta = \sum_{n = 0}^{\infty} \sum_{k = 0}^{\infty} G^{n + 1} \eta^k \Delta_n^{\left(k\right)}. 
\end{equation}
Matching at fixed order in $G$, we can relate orders in the eikonal exponent $\chi$ to the coefficients of $\tilde{A}$ as given by \autoref{eq:Aexpansion}
\begin{subequations}
\begin{align}
\label{eq:chi0amps}
    i\chi _0 &= i\tilde{A}_{0}^{(0)}  ,\\
    i\chi _1 &= i\tilde{A}_{1}^{(0)}- i\tilde{A}_{0}^{(0)} \,i\tilde{A}_{0}^{(1)}. 
\label{eq:chi1amps}
\end{align}
\end{subequations}
Thus, the eikonal exponent at leading eikonal order involves subleading eikonal terms in the expansion of $\tilde{A}$. This is pictorally represented in \autoref{fig:Matching Table} 

\begin{figure}
\begin{center}
\begin{tikzpicture}[scale=1.2]
%lattice: columns N=1..4 at x=1..4 ; rows k=4,3,2,1 at y=3,2,1,0
\foreach \N in {1,...,4} {\node at (\N,3.75) {\footnotesize $G^{\N}$};}
\foreach \k/\y in {-4/3,-3/2,-2/1,-1/0} {\node[anchor=east] at (0.45,\y) {\footnotesize $\eta^{\k}$};}
% E^1 row (y=0) blue
\foreach \N in {1,...,4} {\fill[RoyalBlue!22] (\N-0.47,0-0.47) rectangle (\N+0.47,0+0.47);}
%\node at (-1.5, 0) {semi-classical $\Rightarrow$};
% diagonal E^N: (2,1),(3,2),(4,3) grey; N=1 sits in the E^1 row: hatched over blue
\fill[gray!45] (2-0.47,1-0.47) rectangle (2+0.47,1+0.47);
\fill[gray!45] (3-0.47,2-0.47) rectangle (3+0.47,2+0.47);
\fill[gray!45] (4-0.47,3-0.47) rectangle (4+0.47,3+0.47);
\fill[pattern=north east lines,pattern color=gray!70] (1-0.47,0-0.47) rectangle (1+0.47,0+0.47);
% cell borders
\foreach \N in {1,...,4} {\foreach \y in {0,...,3} {\draw[gray!40] (\N-0.47,\y-0.47) rectangle (\N+0.47,\y+0.47);}}
% labels
\node[RoyalBlue!80!black] at (1,0) {\footnotesize \eqref{eq:chi0amps}};
\node[RoyalBlue!80!black] at (2,0) {\footnotesize \eqref{eq:chi1amps}};
\node[RoyalBlue!80!black] at (3,0) {\footnotesize \autoref{a:Reconstruction All Orders}};
\node[RoyalBlue!80!black] at (4,0) {\footnotesize \autoref{a:Reconstruction All Orders}};
\node[gray!30!black] at (2,1) {\footnotesize $\tfrac12\chi_0^2$};
\node[gray!30!black] at (3,2) {\footnotesize $\tfrac16\chi_0^3$};
\node[gray!30!black] at (4,3) {\footnotesize $\tfrac1{24}\chi_0^4$};
\draw[line width=0.3mm, decorate, decoration={brace, amplitude=0.4cm}, RoyalBlue] (4.5, -0.6) -- node[anchor=north, yshift=-0.4cm] {\small semi-classical \eqref{eq:Full Hawking A}} (0.5, -0.6);
\end{tikzpicture}
\end{center}
\caption{Matching Phase to Amplitude. This shows the double expansion of $1+i\tilde{A}$ by order in $G$ and $\eta$. In this case, we are interested in the eikonal exponent, $\chi$, given by the logarithm of this expansion, corresponding to the $\eta^{-1}$ order and highlighted in blue. On the other hand, the leading eikonal of $\tilde{A}$ contributes to the terms along the diagonal and hence contribute to powers of $\chi_0$, the leading weak field eikonal exponent.}
\label{fig:Matching Table}
\end{figure}

The amplitudes $\tilde{A}_1^{\left(0\right)}$ and $\tilde{A}_0^{\left(1\right)}$ can be computed using tensor manipulation packages such as \verb!xAct! \cite{xTensor}; refer to \autoref{a:SubleadingComps} for the details of this computation. For 4d Reissner--Nordstr\"{o}m, one finds
\begin{subequations}
\begin{align}
    i\tilde{A}^{(1)}_0 &= -\frac{2r_s}{v} - \frac{4r_q^2}{v^2}\, , \\
    i\tilde{A}^{(0)}_1 &= \left(\frac{r_s^2 }{4v} +\frac{r_s r_q^2}{2v^2}\right) 16 iE\log \left(\frac{-v}{4 E \lambda_{\infty}}\right) -\frac{4 iE r_s^2}{v} -\frac{16 iE r_s r_q^2}{v^2} -\frac{64 iE r_q^4}{3 v^3}.
\end{align}
\end{subequations}
Note that these next to leading order terms do not appear to adjust the coefficient of $\log\left(-v/\ell\right)$ in \autoref{eq:leadingEikExp}. Indeed, at any fixed order in the expansion, it will not be apparent that the log changes. 

Substituting these coefficients in the expansion of the amplitudes into \eqref{eq:chi1amps}, we find
\begin{equation}
    i\chi_1 = -\frac{4i E r_s^2}{v}-\frac{8iE r_s r_q^2}{v^2}- \frac{16i E r_q^4}{3v^3} \, .\label{eq:chi1computed}
\end{equation}
We now make the observation that the sub-leading eikonal computation is indeed the expansion under the integrand in $h$ of \autoref{eq:Ray Tracing from Amplitudes}: 
\begin{subequations}
\begin{align}
    i\chi_0 &= 2iE\int_{-v/2}^{\infty} \dd{r} h(r),\\
    i\chi_1 &= 2iE\int_{-v/2}^{\infty} \dd{r} h^2(r),
\end{align}
\end{subequations} 
where we continue to drop the contributions to the Coulomb phase coming from divergences at the upper bound of the $r$ integral. This pattern persists to all orders as demonstrated in \autoref{a:Reconstruction All Orders} for any Kerr--Schild $h(r)$, 
\begin{equation}
\label{eq:chi n-1}
i\chi_{n - 1} = 2i E \int_{-v/2}^{\infty} \dd{r} h^{n}\left(r\right).
\end{equation}
Notice that this leads to a geometric series: if all $\chi_n$ are summed over, we recover the leading eikonal approximation in \autoref{eq:Ray Tracing from Amplitudes}, despite the fact that it is done order by order in $G$. 

As a result, to recover the same coefficient of $\log\left(-v/\ell\right)$ as in \autoref{eq:4dRN}, we would need to go to \textit{all} orders in the weak field \textit{and} eikonal expansions for $\tilde{A}$. As we will show in the next section, and as has been known from different arguments e.g. in Ref.~\cite{Parker:2009}, the coefficient of the log corresponds to Hawking temperature and so this implies that one cannot extract the Hawking temperature at any fixed order in the expansion of $\tilde{A}$. This also offers a perspective on why exponentiating before resummation does not yield the correct temperature. The logarithm comes from the $1/r$ integrand in the $r$ integral. For the full result in \autoref{eq:Ray Tracing from Amplitudes}, this corresponds to a sum of the residues over all zeroes in $1 - h$. This also, however, corresponds to the singularity at the boundary of the convergence in the sum over $h^n$. In order to see all the poles from $1/\left(1 - h\right)$, one would need to carry through full resummation of the insertions of $h$.  

\section{The Stokes Line of Bogoliubov Coefficients}
\label{s:StokesLine}
For generic $h\left(r\right)$, the integral in \autoref{eq:Ray Tracing from Amplitudes} cannot be computed exactly. In light of this, a number of methods for extracting the relevant physical information from the Bogoliubov coefficient have been proposed. For example, as illustrated in \cite[Ch.~4]{Parker:2009}, one can compute the ratio of the Bogoliubov $\alpha$ coefficient to the Bogoliubov $\beta$ coefficient. In the case of a 4d Hawking black hole, at least, this turns out to be an exponential factor. Then, using the unitarity relationships $\abs{\alpha}^2 - \abs{\beta}^2 = 1$, one can deduce $\abs{\beta}^2$ and find the appropriate Hawking spectrum.

In this section, we offer at least one alternative way of extracting the temperature by studying the \textit{analytic} properties of the Hawking amplitude. More precisely, we will focus on the generalized amplitude, $\mathcal{B}$, which is related to $\mathcal{A}$ by analytically continuing $E$ to $-E$ \cite{Aoude:2026bek}:
\begin{equation}
\mathcal{B} = \frac{2\pi}{E_0} \int \dd{v} \e^{-i E_0 v} \exp\left[-i E \left( v + \int_{-v/2}^{\infty} \dd r' \frac{2h\left(r^\prime\right)}{1 - h\left(r^\prime\right)} \right)\right].
\end{equation}
Recall that the lower bound of the $r^\prime$ integral corresponds to last rays which do not escape the horizon. While $\mathcal{A}$, which is proportional to the Bogoliubov $\alpha$ coefficient, was computed from the one-to-one transition amplitude, this generalized amplitude, $\mathcal{B}$, which corresponds to the Bogoliubov $\beta$ coefficient, can be computed from a zero-to-two particle production amplitude. We will show that upon analytically continuing the Bogoliubov coefficient to the complex energy plane, there is a \textit{Stokes line} located at $\text{Im}\left[E\right] = 2\pi T_H$. This means that when evaluating $\mathcal{B}$ one gets a different answer for energies with imaginary part greater than $2\pi T_H$ than for energies with imaginary part less than $2\pi T_H$. This will show that the full answer has a pole located at $E = 2\pi i n T_H$ for any positive integer $n$. The existence and location of these poles is a signature of thermality. We reserve additional examples, including a black hole surrounded by a Hernquist dark matter profile and a static observer a finite distance from a Schwarzschild black hole, for \autoref{a:AdditionalExamples}. For a helpful review of saddle point methods and Picard--Lefschetz theory we use in this section, see \cite{Witten:2010cx, Dunne:2025mye}.

\subsection{Four-Dimensional Schwarzschild}
To illustrate the main points of this section, we will focus on the well-studied case of a 4d Schwarzschild black hole, where $h\left(r\right) = r_s/r$. The generalized amplitude is then given by \autoref{eq:4dSchw Result}
\begin{equation}
\mathcal{B} = \frac{2\pi}{E_0} \int_{-\infty}^{-2r_s} \dd{v} \e^{S\left(v\right)}, 
\end{equation}
where $S\left(v\right)$ is the action given by 
\begin{equation}
S\left(v\right) = -i \left(E + E_0\right) v + 2i r_s E \log\left(\frac{v + 2 r_s}{-\ell}\right). 
\end{equation}
Here, $\ell$ is a length scale used to regulate the divergence from the upper bound of the $r$ integral. The remaining divergent piece is a constant phase, called the Coulomb phase, that will cancel out from the physical observable $\abs{\mathcal{B}}^2$. 

\begin{figure}
\begin{center}
\includegraphics[scale=0.95]{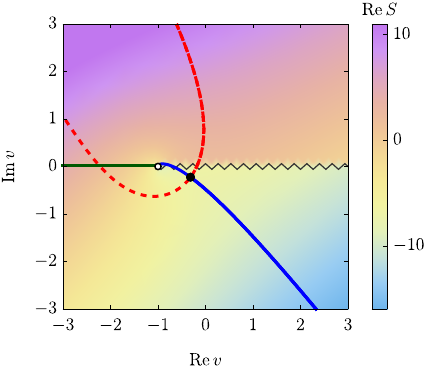} \quad \includegraphics[scale=0.95]{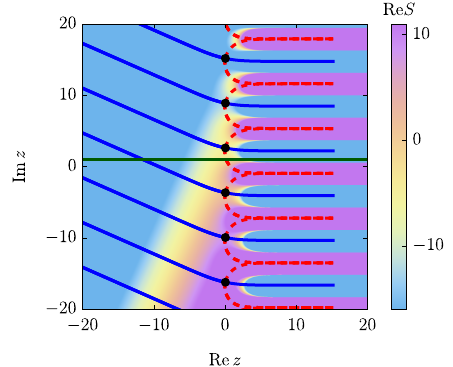}
\end{center}
\caption{Saddles and Contours for 4d Hawking. The saddle points are shown by a black dot, the steepest ascent contours by a dashed red line, the steepest descent contours by a solid blue line, the integration contour by a solid green line, and the branch cut by a zig-zag. In the $v$ variable (left) the steepest ascent and descent contours run into the branch cut and wrap around the saddle point on subsequent Riemann sheets. This is more clearly visible once the branch cut and Riemann sheets are unraveled in the $z$ variable (right). For concreteness, we plot this for $2 r_s E = 1.1 - 1.5 i$.}
\label{fig:4dHawkingThimbles}
\end{figure}

An integral of this form can be studied using a saddle point approximation. For this action, the saddle point and the steepest ascent and descent contours are shown in the left panel of \autoref{fig:4dHawkingThimbles}. As also shown there, the presence of the branch cut in $S\left(v\right)$ forces the contours onto different Riemann sheets. It is therefore helpful to utilize \textit{exponential coordinates} \cite{Farrell:2022mzv} 
\begin{equation}
\operatorname{e}^z = \frac{v + 2 r_s}{-\ell}. 
\end{equation}
Employing this transformation and accounting for the Jacobian factor, we obtain a new expression for the generalized amplitude, 
\begin{equation}
\mathcal{B} = \frac{2\pi \ell}{E_0} \e^{2ir_s\left(E_0 + E\right)} \int_{-\infty}^{\infty} \dd{z} \e^{\tilde{S}\left(z\right)},
\end{equation}
where
\begin{equation}
\tilde{S}\left(z\right) = \left(1 + 2 i E r_s\right)z + i\left(E_0 + E\right)\ell \, \operatorname{e}^z.
\end{equation}
The saddle points for this action are the solutions of $\tilde{S}^\prime\left(z\right) = 0$:
\begin{equation}
\label{eq:Schwarzschild Saddles}
z_n = \log\left(\frac{i - 2 E r_s}{\left(E_0 + E\right)\ell}\right) + 2\pi i n,
\end{equation}
for any integer $n$. In this variable, the action no longer has a branch cut, but the saddle point that was previously there on every Riemann sheet is visible in the $z$ complex plane as shown in the right panel of \autoref{fig:4dHawkingThimbles}. 

\begin{figure}
\centering
\newlength{\insetw}\setlength{\insetw}{0.44\linewidth}
\begin{tikzpicture}[
    >={Stealth[length=2.6mm,width=2.2mm]},
    link/.style={->, line width=1.1pt, black!75},
    saddlelabel/.style={font=\sffamily\small\bfseries},
    insettitle/.style={font=\small, inner sep=2pt},
  ]

  \fill[RoyalBlue,  fill opacity=0.40] (0,0)   rectangle (5,2.8);  
  \fill[BurntOrange, fill opacity=0.55] (0,2.8) rectangle (5,4.2);
  \draw[black,line width=0.6pt] (0,0) rectangle (5,4.2);
  
  \draw[black,line width=1.1pt] (0,2.8) -- (5,2.8);
  \node[anchor=west,font=\small] at (5.15,2.8) {$\operatorname{Im}E = 2\pi T_{H}$};

  \draw[black,line width=0.4pt] (0,1.8) -- (5.05,1.8);
  \node[anchor=west,font=\small] at (5.15,1.8) {$\operatorname{Re}E$};
  \draw[black,line width=0.4pt] (2.5,0) -- (2.5,4.45);
  \node[anchor=south,font=\small] at (2.5,4.48) {$\operatorname{Im}E$};

  \node[saddlelabel] at (2.55,3.55) {$\infty$ relevant saddles};
  \node[saddlelabel] at (1.95,0.85) {1 relevant saddle};

  \node[inner sep=0pt, anchor=north east] (upper) at (2.35,-1.0)
        {\includegraphics[width=\insetw, trim={3bp 0bp 35bp 6bp}, clip]{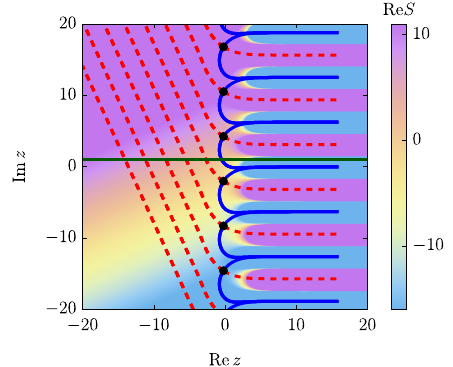}};
  \node[insettitle, anchor=south] (upperT) at (upper.north) {};

  \node[inner sep=0pt, anchor=north west] (lower) at (2.65,-1.0)
        {\includegraphics[width=\insetw, trim={3bp 0bp 35bp 6bp}, clip]{SchLowThimblesPlot.pdf}};
  \node[insettitle, anchor=south] (lowerT) at (lower.north) {};

  \node[inner sep=0pt, anchor=north west] (cbar)
        at ([xshift=0.012\insetw, yshift=0.0335\insetw]lower.north east)
        {\includegraphics[width=0.1732\insetw, trim={183bp 26bp 3bp 0bp}, clip]{SchHighThimblesPlot.pdf}};

  \coordinate (oTap) at (0.7,3.15);
  \fill[black] (oTap) circle (1.6pt);
  \draw[link] (oTap) to[out=180,in=90] (upperT.north);

  \coordinate (bTap) at (4.10,0.70);
  \fill[black] (bTap) circle (1.6pt);
  \draw[link] (bTap) to[out=-20,in=90] (lowerT.north);

\end{tikzpicture}
\caption{Stokes Geometry of Bogoliubov Coefficients. In the top panel, we highlight the universal existence of a Stokes line at $\operatorname{Im}E = 2\pi T_H$. This divides the region in which there is one relevant saddle for the saddle point approximation of $\mathcal{B}$ below the Stokes line and an infinite number above the Stokes line. The bottom panels show the steepest ascent and descent curves in each region for $E = \pm\left(1.1 - 1.5 i\right)$; the color bar applies to both.}
\label{fig:StokesGeometry}
\end{figure}

To explore the analytic structure of the result of the integral, $\mathcal{B}$, as a function of the outgoing energy, $E$, we study the saddle point approximation for different values of the energy $E$. To begin, we start with an energy with slightly negative imaginary part, shown by the lower right black point in the center of \autoref{fig:StokesGeometry}. The saddle points (black points), steepest ascent (red, dashed) and descent (blue, solid) contours, and the integration contour (green) are shown on the bottom right panel of \autoref{fig:StokesGeometry}. In order for the integral to converge, we must then deform the integration contour (green, along real axis) to follow along a steepest descent contour, ensuring exponential fall off at infinity. Then, for the contour to be deformed to lower values of the action, $\tilde{S}$, so that it passes through a saddle point and hence the steepest descent contour, the steepest \textit{ascent} contour must intersect with the integration contour. In other words, the only relevant saddles in the saddle point approximation will be those whose steepest ascent contours pass through the integration contour. In this case, there is only one relevant saddle, located at $n = 0$, slightly above the real axis. In this regime, the generalized amplitude evaluates to
\begin{equation}
\mathcal{B}_{\text{low}} = \frac{2\pi \ell}{E_0} \sqrt{\frac{2\pi}{1 + 2 i r_s E }} \e^{2i r_s E_0 - 1} \left(\frac{i\left(1 + 2 i r_s E\right)}{\left(E_0 + E\right)\ell}\right)^{1 + 2 i r_s E},
\end{equation}
where low means sufficiently small imaginary part of $E$. On the other hand, if we evaluate the same integral for energies with sufficiently large positive imaginary part, we instead have the situation shown in the top left panel of \autoref{fig:StokesGeometry}, where, in this case, all saddles in the lower half plane are relevant and contribute in the saddle point approximation. Hence, for $\operatorname{Re} E < 0$ we have
\begin{subequations}
\begin{align}
\mathcal{B}_{\text{high}} &= \frac{2\pi \ell}{E_0} \sqrt{\frac{2\pi}{1 + 2 i r_s E }} \e^{2i r_s E_0 - 1} \left(\frac{i\left(1 + 2 i r_s E\right)}{\left(E_0 + E\right)\ell}\right)^{1 + 2 i r_s E} \sum_{n = 0}^{\infty} \e^{4 n \pi E r_s},\\
&= \frac{2\pi \ell}{E_0} \sqrt{\frac{2\pi}{1 + 2 i r_s E }} \e^{2i r_sE_0 - 1} \left(\frac{i\left(1 + 2 i r_s E\right)}{\left(E_0 + E\right)\ell }\right)^{1 + 2 i r_s E} \left(\frac{\e^{4\pi E r_s}}{\e^{4\pi E r_s} - 1}\right).
\end{align}
\end{subequations}
where high means sufficiently large imaginary part of $E$. For $\operatorname{Re} E < 0$, we instead have negative $n$ yielding the same result. Notice here that the generalized amplitude has poles for $E = in/\left(2 r_s\right)$. Indeed, this prediction coincides with the poles in the full answer for the Bogoliubov coefficient, which is a $\Gamma$ function, see e.g. \cite[eq. (3.34)]{Aoude:2024sve}. Since this is a Bose-Einstein distribution, we associate these poles with the temperature, 
\begin{equation}
T_H = \frac{1}{4\pi r_s}.
\end{equation}
Indeed, this is the correct prediction for the Hawking temperature.

The observation that the number of contributing saddles in the complex $E$ plane jumps is called a \textit{Stokes phenomenon}. The precise location where this jump occurs is called a \textit{Stokes line}. This can be found analytically. In order for a set of steepest ascent curves to go from eventually crossing an integration contour to not, they eventually must hit another saddle point and change direction -- see for example \cite[Fig.~2]{Witten:2010cx}. By the gradient flow equations that govern the steepest ascent and descent curves, in order for the saddle points to be connected, their imaginary parts must coincide. Therefore, a necessary (but not sufficient) condition for the appearance of a Stokes line is that 
\begin{equation}
\text{Im}\left[S\left(z_n\right)\right] - \text{Im}\left[S\left(z_{n^\prime}\right)\right] = 0
\end{equation}
for two adjacent saddles $z_n$ and $z_{n^\prime}$. Since two saddle points, per \autoref{eq:Schwarzschild Saddles}, are separated by $2\pi i$ \textit{and} $\tilde{S}\left(z\right)$ only depends on $z$ and $e^{z}$, we only need to note that
\begin{equation}
\tilde{S}\left(z + 2\pi i\right) - \tilde{S}\left(z\right) = 2\pi i \left(1 + 2 i E r_s\right).
\end{equation}
This is zero when
\begin{equation}
\operatorname{Im}\left[E\right] = \frac{1}{2r_s} = 2\pi \, T_H, 
\end{equation}
where in the last equality we rewrite this answer in terms of the Hawking temperature. This therefore justifies the identification of the location Stokes line with the Hawking temperature in \autoref{fig:StokesGeometry}.

We note, importantly, that the location of the Stokes line is coordinate dependent. In the original $v$ variable, the Stokes line would have occurred at the real axis, but when switching to exponential coordinates we included an exponential Jacobian factor thereby shifting the Stokes line upward in the energy plane. We have demonstrated, however, the invariant statement, which is that above the Stokes line one sums over an infinite number of saddles and upon doing this sum, we find that the location of the poles in $\mathcal{B}$ begin exactly at $\operatorname{Im} E = 2\pi T_H$. 

\subsection{Higher-Dimensional Black Holes}
We now turn our attention to higher dimensional black holes with results given in \autoref{eq:ST u} and \autoref{eq:RN u}. Since the former case can be recovered from the latter in the limit $r_- \rightarrow 0$ and $r_+ \rightarrow r_s$, we focus on higher-dimensional Reissner--Nordstr\"{o}m black holes. The generalized amplitude we study is
\begin{equation}
\mathcal{B}_{\text{RN}} = \frac{2\pi}{E_0} \int_{-\infty}^{-2r_+} \dd{v} \e^{S_{\text{RN}}\left(v\right)},
\end{equation}
where, using the fact that the sum over the $n^{\text{th}}$ roots of unit is $0$ to simplify \autoref{eq:RN u}, the action is
\begin{equation}
\begin{aligned}
S_{\text{RN}} &= -i \left(E_0 + E\right)v + \left[
\frac{i E}{\kappa_+} \sum_{k = 0}^{d - 4} \omega_k \log\left(1 + \frac{2 r_+}{v} \omega_k\right) - \frac{i E}{\kappa_-}\sum_{k = 0}^{d - 4} \omega_k \log\left(1 + \frac{2r_-}{v} \omega_k\right)
\right], 
\end{aligned}
\end{equation}
where 
\begin{equation}
\omega_k \equiv \e^{\frac{2\pi i k}{d - 3}} \quad \text{and} \quad \kappa_\pm \equiv \frac{\left(d - 3\right)\left(r_+^{d - 3} - r_-^{d - 3}\right)}{2 r_\pm^{d - 2}}.
\end{equation}
Here, $\omega_k$ are the roots of unity while $\kappa_{\pm}$ have the physical interpretation of being the surface gravity on the outer and inner horizons respectively. 

We now transform to exponential coordinates just as in the previous section, with the following variable change
\begin{equation}
\e^z = \frac{v + 2 r_+}{- 2 r_+}.
\end{equation}
Implementing this, we find that the generalized amplitude is now
\begin{equation}
\mathcal{B}_{\text{RN}} = \frac{4\pi r_+}{E_0}\, \e^{2i\left(E_0 + E\right) r_+} \int_{-\infty}^{\infty} \dd{z} \e^{\tilde{S}_{\text{RN}}\left(z\right)},
\end{equation}
where
\begin{equation}
\begin{aligned}
\tilde{S}_{\text{RN}}\left(z\right) = {}& \left(1 + \frac{i E}{\kappa_+}\right) z + 2i\left(E_0 + E\right) r_+ \e^{z} \\
& + \frac{i E}{\kappa_+} \sum_{k = 1}^{d - 4} \omega_k \log\left(1 + \e^{z} - \omega_k\right) - \frac{i E}{\kappa_-} \sum_{k = 0}^{d - 4} \omega_k \log\left(1 + \e^{z} - \frac{r_-}{r_+}\, \omega_k\right),
\end{aligned}
\end{equation}
Once again, the saddle points are solutions of $\tilde{S}_{\text{RN}}^\prime\left(z\right) = 0$. Compared to the case of a four-dimensional Hawking black hole, there are $2\left(d - 3\right)$ saddle points per Riemann sheet and even after transforming to exponential coordinates, there are remaining branch cuts.

\begin{figure}
\centering
\ifdefined\insetw\else\newlength{\insetw}\fi
\setlength{\insetw}{0.44\linewidth}
\begin{tikzpicture}[
    >={Stealth[length=2.6mm,width=2.2mm]},
    link/.style={->, line width=1.1pt, black!75},
    stokesray/.style={dashed, black!75, line width=0.8pt},
  ]

  \coordinate (Crit) at (2.5,2.8);

  \begin{scope}
    \clip (0,0) rectangle (5,4.2);
    \fill[RoyalBlue,   fill opacity=0.40] (0,0)   rectangle (5,2.8);  
    \fill[BurntOrange, fill opacity=0.55] (0,2.8) rectangle (5,4.2); 
    \foreach \a in {150,110,70,30,-150,-110,-70,-30}
      \draw[stokesray] (Crit) -- +(\a:4cm);
  \end{scope}

  \draw[black,line width=0.6pt] (0,0) rectangle (5,4.2);

  \draw[black,line width=1.1pt] (0,2.8) -- (5,2.8);
  \node[anchor=west,font=\small] at (5.15,2.8) {$\operatorname{Im}E = 2\pi T_{H}$};

  \draw[black,line width=0.4pt] (0,1.8) -- (5.05,1.8);
  \node[anchor=west,font=\small] at (5.15,1.8) {$\operatorname{Re}E$};
  \draw[black,line width=0.4pt] (2.5,0) -- (2.5,4.45);
  \node[anchor=south,font=\small] at (2.5,4.48) {$\operatorname{Im}E$};

  \fill[black] (Crit) circle (1.6pt);

  \node[inner sep=0pt, anchor=north east] (upper) at (2.35,-0.8)
        {\includegraphics[width=\insetw, trim={3bp 0bp 35bp 6bp}, clip]{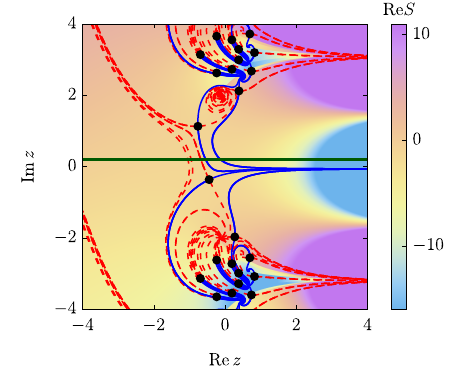}};
  \node[inner sep=0pt, anchor=north west] (lower) at (2.65,-0.8)
        {\includegraphics[width=\insetw, trim={3bp 0bp 35bp 6bp}, clip]{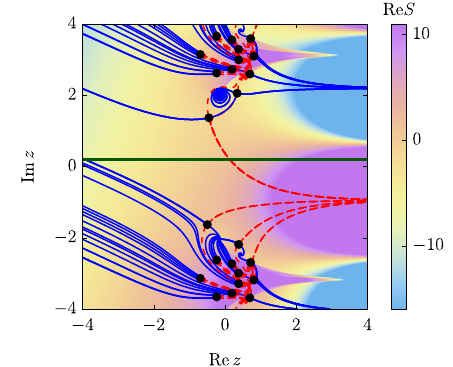}};

  \node[inner sep=0pt, anchor=north west] (cbar)
        at ([xshift=0.012\insetw, yshift=0.0335\insetw]lower.north east)
        {\includegraphics[width=0.1732\insetw, trim={183bp 26bp 3bp 0bp}, clip]{RNLowThimblesPlot.pdf}};

  \coordinate (oTap) at (0.7,3.15);
  \fill[black] (oTap) circle (1.6pt);
  \draw[link] (oTap) to[out=180,in=90] ([yshift=1mm]upper.north);

  \coordinate (bTap) at (4.10,0.70);
  \fill[black] (bTap) circle (1.6pt);
  \draw[link] (bTap) to[out=-20,in=90] ([yshift=1mm]lower.north);

\end{tikzpicture}
\caption{Stokes Geometry of Bogoliubov Coefficients for Reissner--Nordstr\"{o}m black holes. In the top panel, the solid line at $\operatorname{Im}E = 2\pi T_H$ separates a finite number of relevant saddles in the saddle point approximation of $\mathcal{B}_{\text{RN}}$ below it from an infinite number above it. The dashed lines are the additional Stokes rays, across which the number of relevant saddles changes by a finite amount. The bottom panels show the steepest ascent (red, dashed) and descent (blue, solid) contours in the $z$ variable for $d = 10$, for representative energies in each region; the color bar applies to both. Compared to the case of a 4d Schwarzschild black hole, there are remaining branch cuts after switching to exponential coordinates.}
\label{fig:StokesGeoRN}
\end{figure}

Due to the presence of additional saddles per Riemann sheet, there are new Stokes rays resulting from the points where the number of steepest ascent contours that pass through the integration contour changes. However, one of these Stokes lines is unique -- where the number of relevant saddles goes from finite to infinite. As before, this is the Stokes line associated to the Hawking temperature. The Stokes geometry in this case is shown in \autoref{fig:StokesGeoRN}. 

We would like to compute these Stokes lines analytically. As discussed in the previous section, the temperature Stokes line occurs whenever a steepest descent curve connects saddles across different Riemann sheets. To this end, we note that
\begin{equation}
\tilde{S}_{\text{RN}}\left(z + 2\pi i\right) - \tilde{S}_{\text{RN}}\left(z\right) = 2\pi i \left(1 + \frac{i E}{\kappa_+}\right).
\end{equation}
In particular, including when evaluated on saddles, the imaginary part of this difference vanishes whenever
\begin{equation}
\text{Im}\left(E\right) = \kappa_+.
\end{equation}
Identifying this with $2\pi T_H$, we find that the horizontal Stokes line differentiating a finite from an infinite number of relevant saddles implies the Hawking temperature is 
\begin{equation}
\label{eq:RNtemp}
T_H = \frac{\kappa_+}{2\pi},
\end{equation}
which is the known Hawking temperature for a Reissner--Nordstr\"{o}m black hole. The additional Stokes lines indicate jumps either from a finite to finite number or infinite to infinite number of saddles.

To demonstrate the effect of the Stokes line on the evaluation of the Bogoliubov coefficient, we denote the saddles as
\begin{equation}
z_\ast = z_j + 2 \pi i n,
\end{equation}
where $z_j$ denotes the saddles on any given individual Riemann sheet. Below the saddle, only a finite number of saddles are relevant -- in particular those on one Riemann sheet, given by $n = 0$. Hence, 
\begin{equation}
\mathcal{B}_{\text{RN}}^{\text{low}} = \sum_{j} \frac{4\pi r_+}{E_0} \e^{2i\left(E_0 + E\right)r_+}\sqrt{\frac{2\pi}{-\tilde{S}^{\prime \prime}_{\text{RN}}\left(z_j\right)}} \e^{\tilde{S}_{\text{RN}}\left(z_j\right)},
\end{equation}
where the low superscript indicates that this is the evaluation below the temperature Stokes line. Above that line, there is an additional contribution
\begin{subequations}
\begin{align}
\mathcal{B}_{\text{RN}}^{\text{high}} &= \mathcal{B}_{\text{RN}}^{\text{low}} \sum_{n = 0}^{\infty} \e^{2\pi E n /\kappa_+},\\
&= \mathcal{B}_{\text{RN}}^{\text{low}} \left(\frac{\e^{2\pi E/\kappa_+}}{\e^{2\pi E/\kappa_+} - 1}\right).
\end{align}
\end{subequations}
Once again, as expected, this indicates that the resulting amplitude will have poles at $E = in \kappa_+$, exactly coinciding with the expected poles in a Bose--Einstein distribution with temperature given by \autoref{eq:RNtemp}. All that changes, as you cross into other Stokes regions, is the numerator of this distribution. 

\section{Discussion}
\label{s:Discussion}
As shown in \cite{Aoude:2024sve}, the Hawking spectrum of a black hole can be computed from a generalized scattering amplitude, called the Hawking amplitude. In this paper, we compute these generalized amplitudes in the semi-classical limit for spherically symmetric black holes whose geometries admit a Kerr--Schild form, including Schwarzschild--Tangherlini and Reissner--Nordstr\"{o}m black holes in $d$ dimensions. The authors of \cite{Aoude:2024sve} raised an interesting question as to whether the derivation of a thermal spectrum with the correct temperature for a 4d Schwarzschild black hole by exponentiating only a tree-level diagram was a coincidence or not. Our analysis shows that this is indeed special to 4d Schwarzschild and we illustrate how this coincidence arises in the simpler setting of acoustic waves in \autoref{s:AcousticRadiation}. There, we also show that the weak field and eikonal limits do not generically commute. Indeed, in \autoref{s:HawkingSpectrum}, we compute the Hawking amplitude solely using a geometric optics approximation and observe that the location of its branch cuts and coefficients of logarithms are different from a Hawking amplitude computed using the weak field expansion before the eikonal approximation. This has a physical origin -- the rays that contribute most to Hawking radiation are the ones in which $h/\eta \gg 1$, where $h$ parametrizes the weak field expansion and $\eta$ parametrizes the eikonal approximation.

The evaluation of the Hawking amplitudes for arbitrary background is not generically possible. We instead opt to study these integrals using a saddle point analysis in \autoref{s:StokesLine}. We show that there is a line in the complex plane of the outgoing energy, located at $\operatorname{Im} E = 2\pi T_H$, where the number of saddle points that need to be included in the saddle point approximation jumps from a finite to infinite number. We propose that this Stokes line is universal for observers outside a black hole and hence provides a novel way of extracting the Hawking temperature without having to perform the integral. More concretely, the resummation of an infinite number of saddles above the Stokes line is sufficient to show that the Hawking amplitude has poles located at $E = 2\pi i n T_H$, as expected for a thermal distribution. More complicated black hole backgrounds, such as $d$-dimensional Reissner--Nordstr\"{o}m, have additional Stokes rays, but only one line associated to the temperature results in a jump from a finite to an infinite number of relevant saddles. 

It is interesting to note that many alternative methods for extracting the temperature from the Bogoliubov coefficient without evaluating the integral rely on unitarity. For example, one can show that if the Bogoliubov coefficients are related by an exponential factor, then the unitarity relationship $\abs{\alpha}^2 - \abs{\beta}^2 = 1$ is enough to compute the thermal spectrum \cite{Parker:2009}. Alternatively, the authors of Ref.~\cite{Akpinar:2026oni} demonstrated that the temperature of a black hole can be extracted by saturating the unitarity bound that the probability of absorbing a particle can never exceed one. Our approach, on the other hand, seems to extract the same information without using unitarity in any obvious way. It would be interesting to investigate whether or not the universal Stokes line itself is a consequence of unitarity. Such an investigation would also be of interest since the methods we use to compute the amplitude in this work are manifestly off-shell, and so it would be useful to reformulate the details of the derivation in a manifestly on-shell way and one way to do so may be to explicitly incorporate unitarity. 

Knowledge of the Stokes geometry has many uses beyond evaluating the analytic structure of the integral, as we have done in this paper. For example, the authors of Ref.~\cite{Miyachi:2025ptm} derived an exact formula for the quasinormal modes of a black hole by studying how solutions of the Regge--Wheeler equations transform across Stokes lines. It would therefore be worthwhile to explore the crossing of these Stokes lines using a similar approach.

Other natural extensions include extending our computation to more generic black holes and observers. For example, a Kerr black hole breaks spherical symmetry and hence leads to a more involved computation. More recently, there has been a growing interest in Hawking radiation measured by different observers, such as a static or free-falling observer \cite{Barcelo:2010pj, Barcelo:2010xk, Barbado:2011dx, McMaken:2024}. In \autoref{a:AdditionalExamples}, we showed that the Bogoliubov coefficient measured by a static observer a finite distance from the Schwarzschild black hole still has the same universal Stokes line, now at the temperature measured by the static observer. However, for this, we used known ray-tracing relationships. It would be interesting to explore how to compute the one-to-one transition amplitude as seen by different observers.

\acknowledgments

We thank Wayne Ariston, Andrea Casale, Craig Clark, Alice Garoffolo, Matteo Sergola, and Leonardo Stanzani for useful comments and discussions. O.L. is supported by the Leinweber Institute for Theoretical Physics at the University of Michigan and the LB Science Summer Research Fellowship of Columbia University. D.M. is supported by the US Department of Energy under award number DE-SC0024224, the Sloan Foundation, and the Mani L. Bhaumik Institute for Theoretical Physics. S.M. is supported by the U.S. Department of Energy, Office of Science, under Award Number DE-SC0011941. J.S. is supported by funds provided by the Center for Particle Cosmology at the University of Pennsylvania.

\appendix

\section{Bogoliubov Coefficients from Ray Tracing}
\label{a:Bogoliubov Ray Tracing}

In this appendix, we will review the computation of the Bogoliubov coefficients from the ray tracing argument originally proposed by Hawking \cite{Hawking:1974rv}. We do this for a spherically symmetric metric cast in Kerr--Schild form, as in \autoref{eq:Kerr-Schild Form}. The methods we use here can be reviewed in \cite[Chapter 4]{Parker:2009}.

\begin{figure}
\begin{center}
\begin{tikzpicture}
\filldraw[BrickRed!20, opacity=0.5] (0, 1.5) decorate[decoration={zigzag, segment length=0.2cm}] {-- (1.5, 1.5)} -- (2.25, 0.75) -- (1.5, 0) -- (0, 1.5);
\filldraw[RoyalBlue!20, opacity=0.5] (0, 1.5) -- (0, -1.5) -- (1.5, 0) -- (0, 1.5);
\draw[line width=0.3mm] (0, -1.5) -- (0, 1.5);
\draw[line width=0.3mm, decorate, decoration={zigzag, segment length=0.2cm}] (0, 1.5) -- (1.5, 1.5);
\draw[line width=0.3mm] (0, -1.5) -- node[midway, below right] {$\mathcal{I}^-$} (2.25, 0.75) -- node[midway, above right] {$\mathcal{I}^+$} (1.5, 1.5);
\node at (1.5, 1.75) {$i^+$};
\node at (0, -1.75) {$i^-$};
\node at (2.5, 0.75) {$i^0$};
\draw[line width=0.2mm, dashed] (0, 1.5) -- node[midway, above, rotate=-45, xshift=0.2cm] {$V = 0$} (1.5, 0);
\draw[line width=0.3mm, decorate, decoration={snake, segment length=0.22cm}, -{Stealth[scale=0.75]}, BurntOrange] (0.5, -1) -- (0.1, -0.6);
\draw[line width=0.3mm, decorate, decoration={snake, segment length=0.22cm}, -{Stealth[scale=0.75]}, BurntOrange] (1, -0.5) -- (0.6, -0.1);
\draw[line width=0.3mm, decorate, decoration={snake, segment length=0.22cm}, -{Stealth[scale=0.75]}, BurntOrange] (2, 0.5) -- (1.6, 0.9);
\node[rotate=90] at (-0.2, 0) {$r = 0$};
\node at (0.75, 1.8) {$r = 0$};
\end{tikzpicture}
\caption{Penrose Diagram. In a collapse background, the blue region corresponds to a flat spacetime, where incoming rays can pass through $r = 0$ and escape to $\infty$. After $V = 0$, the spacetime is Schwarzschild and the infalling rays instead do not escape the horizon and fall towards the singularity. Past and future null infinity are given by $\mathcal{I}^-$ and $\mathcal{I}^+$ respectively. Meanwhile, $i^-$ and $i^+$ are past and future timelike infinity respectively. Then, $i^0$ is spatial infinity. }
\label{fig:PenroseDiagram}
\end{center}
\end{figure}
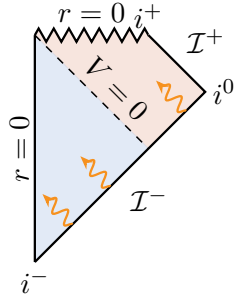

To begin, we note that a Klein--Gordon field admits the following mode decompositions at all times:
\begin{subequations}
\begin{align}
\phi\left(x\right) &= \int \left(a_k P\left(x, k\right) + a_k^\dagger P^\ast\left(x, k\right)\right)\dd{\Phi}\left(k\right)\\
&= \int \left(b_k F\left(x, k\right) + b_k^\dagger F^\ast\left(x, k\right)\right) \dd{\Phi}\left(k\right),
\end{align}
\end{subequations}
where $P$ and $F$ are the past and future mode functions, defined as coinciding with the Minkowski mode functions on past and future null infinity respectively. The Bogoliubov coefficients measure how different the past and future mode functions and hence past and future vacua are. In particular, the first Bogoliubov coefficient is
\begin{equation}
\alpha\left(k, p\right) = \left(F\left(x, p\right), P\left(x, k\right)\right) \equiv \int \left(F^\ast\!\left(x, p\right) i \partial_t P\left(x, k\right) - P\left(x, k\right) i\partial_t F^\ast\!\left(x, p\right)\right) \dd[d-1]{x}. 
\end{equation}
We consider the case of spherically symmetric radiation described by the energy at past null infinity, $E_0$, and the energy at future null infinity, $E$. Hence, the mode functions are given by
\begin{equation}
F\left(u, E\right) = \frac{1}{2Er^{d/2 -1}} \e^{-i E u} \quad \text{and} \quad P\left(v, E_0\right) = \frac{1}{2E_0 r^{d/2 - 1}} \e^{-i E_0 v},
\end{equation}
where $u$ and $v$ are the null coordinates. Plugging this in, we find that the Bogoliubov coefficient is
\begin{equation}
\label{eq:Bogoliubov from Overlap}
\alpha\left(E_0, E\right) = \frac{2\pi}{E} \int_{-\infty}^{v_0} \e^{-iE_0 v} \e^{i E u\left(v\right)} \dd{v}.
\end{equation}
To explain the integration bounds, we look at the collapse background for a Schwarzschild black hole, also called a Vaidya black hole in \autoref{fig:PenroseDiagram}. Not all of the null rays at a position $v$ on past null infinity make it to future null infinity. Those rays, given by $u\left(v\right)$ are the ones that made it to future null infinity without falling into the horizon. The last rays that make it to future null infinity occur at a coordinate $v_0$, which will depend on the specific background in question, but for a 4d Schwarzschild black hole is given by $-2r_s$. 

Up to a factor of $E/E_0$, \autoref{eq:Bogoliubov from Overlap} is the same as the Hawking amplitude in \autoref{eq:Bogoliubov from Amplitude}. The difference is that we must now find $u\left(v\right)$, which amounts to tracing the null geodesic starting from a position $v$ on past null infinity and ending on future null infinity at $u\left(v\right)$. To find the null geodesic, we study the path in \autoref{fig:PenroseDiagram}. To clarify notation, we will use $v$ to label past null infinity, $u$ to label future null infinity while capital $V$ and $U$ will label the bulk null coordinates. 

A ray starting from a position $v$ on past null infinity falls inward towards $r = 0$ along constant $V$ in flat spacetime. It reflects on the boundary and travels along constant $U$. Since at $r = 0$, $V$ and $U$ agree, $v = U$. As it radiates outward along $U \equiv V - 2r$, by definition, it will eventually hit $V = 0$, which defines the moment the black hole is formed. At this point, we will leave flat spacetime and go towards a curved spacetime described by \autoref{eq:Kerr-Schild Form}. There, $v = U = - 2r$. Since it is a null geodesic, in curved spacetime it satisfies
\begin{equation}
\dd{s}^2 = 0 = \left(-1 + h\left(r\right)\right)\dd{V}^2 + 2 \dd{V} \dd{r},
\end{equation}
where we consider a radially outgoing ray, hence $\dd{\theta} = \dd{\varphi} = 0$. Hence, in the curved portion of the Penrose diagram,
\begin{equation}
V\left(r\right) = \int_{-v/2}^{r} \frac{2 \dd{r^\prime}}{1 - h\left(r^\prime\right)} = v + 2r + \int_{-v/2}^{r} \frac{2h\left(r^\prime\right)}{1 - h\left(r^\prime\right)} \dd{r^\prime}. 
\end{equation}
Translating this back to $U = V - 2r$, we see that the $2r$ terms cancel. We can then take the upper bound of the integral to $\infty$ to reach future null infinity giving us
\begin{equation}
\label{eq:Ray Tracing Final}
u\left(v\right) = v + \int_{-v/2}^{\infty} \frac{2h\left(r^\prime\right)}{1 - h\left(r^\prime\right)} \dd{r^\prime}.
\end{equation}
We note that in expressions of the ray tracing relationship, it is helpful to define the location on past null infinity, $v_0$, where the last null ray reaches future null infinity. For $v > v_0$, the rays instead fall into the horizon, once again justifying the upper bound of \autoref{eq:Bogoliubov from Overlap}. 

\section{Computation of the Subleading Eikonal}
\label{a:SubleadingComps}

Computing $\chi_1$ in \eqref{eq:chi1computed} required the amplitude coefficients $\tilde{A}_{1}^{(0)}$ and $\tilde{A}_{0}^{(1)}$. Computing $\tilde{A}_{0}^{(1)}$ follows from isolating the subleading in $\eta$ part of
\begin{equation}
    i\tilde{A}_0(b)=i\int \dd{\lambda} \, p_\mu \sum_{n=0}^\infty \frac{(i\lambda)^n}{n!} \left[(p_\nu - i\partial_\nu )\Box^n h^{\mu\nu}(x) \right]\Big|_{x=b+2\lambda p},
\end{equation}
where $\Box = \partial_\mu \partial^\mu$.

In order to compute $\tilde{A}^{(0)}_1$, the procedure used for the tree amplitudes may be extended to the diagram involving two vertex insertions, indicated by the subscript $1$. The relevant integral is now effectively over two ray-tracing parameters $\lambda_1$ and $\lambda_2$:
\begin{equation}
    i\tilde{A}_1(b) = \frac{i^2}{2} \int \left[\dd{\lambda_1} \frac{\dd[4]{\ell}}{\left(2\pi\right)^4} \e^{-i(b+2\lambda_1 p)\cdot \ell_1}\right]\left[\dd{\lambda}_2 \frac{\dd[4]{\ell_2}}{\left(2\pi\right)^4} \e^{-i(b+2\lambda_2 p)\cdot \ell_2}\right]\, \mathcal{J}(\ell_1, \lambda_1, \ell_2, \lambda_2),
\end{equation}
where $\mathcal{J}$ here is separated into coefficients in the geometric optics ($\eta$) expansion:
\begin{subequations}
\begin{align}
\mathcal{J} &= \mathcal{J}^{(-1)} + \mathcal{J}^{(0)} + \mathcal{J}^{(1)} + \cdots \\
\mathcal{J}^{(-1)}(\ell_1, \ell_2) &\equiv (p \cdot \mathsf{h}(\ell_1) \cdot p)\,(p \cdot \mathsf{h}(\ell_2) \cdot p).
\end{align}
\end{subequations}
Note that we will only need $\mathcal{J}^{\left(0\right)}$ to compute $i\tilde{A}_1^{\left(0\right)}$. After writing the Fourier transforms of $h$ explicitly, integration by parts allows us to again reduce the exponent to a $\delta$-distribution which enforces $x_i = b+2\lambda_i p$. We are then simply left with the integrals over $\lambda_1$ and $\lambda_2$ of
\begin{equation}
\begin{aligned}
\mathcal{J}^{(0)} &= -i\lambda_1 (p\cdot h(x_2)\cdot p)(p^\mu p^\nu \Box_1 h_{\mu\nu}(x_1)) -i\lambda_2 (p\cdot h(x_1)\cdot p)(p^\mu p^\nu \Box_2 h_{\mu\nu}(x_2)) \\
&\qquad -i\left[(p^\mu p^\nu h_{\mu\nu}(x_2)p^\rho \partial_1^\sigma h_{\rho\sigma}(x_1) )+ (p^\mu p^\nu h_{\mu\nu}(x_1)p^\rho \partial_2^\sigma h_{\rho\sigma}(x_2) )\right] \\
&\qquad+2 \Theta(\lambda_1 - \lambda_2) \Big[-i\lambda_2 p^\mu p^\nu p^\alpha p^\beta (\partial_{1,\lambda} h_{\mu\nu}(x_1))( \partial_{2}^\lambda h_{\alpha \beta}(x_2))  \\
&\qquad-i p^\mu p^\nu p^{\alpha} h_{\lambda \alpha}(x_2) \partial_1^{\lambda}h_{\mu\nu} (x_1)\Big] +(1\leftrightarrow 2),
\end{aligned}
\end{equation}
where all derivatives are labeled 1 or 2 according to the $\ell_i$ they correspond to. Computing these integrals for Reissner--Nordstr\"om yields:
\begin{equation}
    i\tilde{A}^{(0)}_1 = \left(\frac{r_s^2 }{4v} +\frac{r_s r_q^2}{2v^2}\right) 16 iE\log \left(\frac{-v}{4 E \lambda_{\infty}}\right) -\frac{4 iE r_s^2}{v} -\frac{16 iE r_s r_q^2}{v^2} -\frac{64 iE r_q^4}{3 v^3}\, .
\end{equation}
Here, $\lambda_{\infty}$ is once again a regulator. Together with $\tilde{A}_{0}^{(1)}$ and $\tilde{A}_{0}^{(0)}$, these compute the subleading eikonal $\chi_1$.

\section{Reconstruction at All Orders}
\label{a:Reconstruction All Orders}

In \autoref{s:AcousticRadiation} we showed that taking the eikonal limit of the leading interaction diagram resulted in an exponentiation that led to the wrong Hawking temperature. We then used a WKB approach in \autoref{s:HawkingSpectrum} to show that the correct $\eta^{-1}$ phase of the amplitude can be computed by working non-perturbatively in $G$. At the end of \autoref{s:HawkingSpectrum}, we showed to $\mathcal{O}(G^2)$ that one can still work perturbatively in $G$ and reconstruct the correct phase. In this appendix, we close the loop and show that this result persists at every order in the weak-field expansion, so that the exponentiation of \autoref{eq:eikExp} holds with $\Delta=\mathcal{O}\left(\eta^{0}\right)$ at every order in $G$, and the sum of the $\eta^{-1}$ row is
\begin{equation}
\label{eq:AllOrdersClaim}
i\chi=2iE\int_{-v/2}^{\infty}\frac{h\left(r\right)}{1-h\left(r\right)}\dd{r},
\end{equation}
the correct exponent of \autoref{eq:Ray Tracing from Amplitudes}. In doing so, we compute a Born series expansion in $G$ with a diagrammatic interpretation, and highlight what eikonalizing the leading interaction diagram drops that resulted in the incorrect temperature.

Along the ray, the eikonal parameter at radius $r$ is $\eta\sim1/\left(Er\right)$, so at fixed order in $G$, a term of order $\eta^{-k}$ is a term of order $E^{k}$ at fixed $r$. We grade all quantities below by their power of $E$ in this sense: a quantity has order $\eta^{-k}$ if it is $\mathcal{O}\left(E^{k}\right)$ as $E\to\infty$ at each fixed order in $G$. Each factor of $h$ or its derivatives carries one power of $G$. We treat every $h\left(r\right)$ term in the equation of motion as a perturbation of flat space, at each order $G^{n}$ keep the exact dependence on $E$ and extract the order-$\eta^{-1}$ part, and sum over $n$ last. The object we solve for is the ratio $\mathcal{F}$ of the full solution to the flat one, obtained from a Born series.

\subsection{The Envelope and its Born Series}
\label{a:Recon Setup}

The full equation of motion for a scalar field in the collapse background of \autoref{eq:Kerr-Schild Form} is
\begin{equation}
\label{eq:CollapseEOM}
\left(\partial_{t}^{2}-\nabla^{2}\right)\phi+\left(\partial_{t}-\partial_{r}-\frac{d-2}{r}\right)\left[h\left(r\right)\Theta\left(V\right)\left(\partial_{t}-\partial_{r}\right)\phi\right]=0 ,
\end{equation}
where $V=t+r$ is the advanced time of \autoref{a:Bogoliubov Ray Tracing}. Along the outgoing ray $t=v+r$ the shell function is $\Theta\left(V\right)=\Theta\left(r+v/2\right)$, so the shell enters every integral below as its lower limit $-v/2$, as in \autoref{eq:Ray Tracing from Amplitudes}. At fixed frequency $E$, the outgoing solution of the free ($h=0$) radial equation is
\begin{equation}
\label{eq:FreeSolution}
\phi_{\text{free}}\left(t,r\right)\equiv r^{-a}\e^{-iEt}\varphi_{+}\left(r\right),
\end{equation}
where $a\equiv\frac{d-2}{2}$ and $\varphi_{+}$ is the outgoing Riccati--Hankel function, the solution of
\begin{equation}
\label{eq:RadialFree}
\varphi^{\prime\prime}+\left(E^{2}-a\left(a-1\right)/r^{2}\right)\varphi=0
\end{equation}
that behaves as $\varphi_{+}\left(r\right)=\e^{iEr}\left(1+\mathcal{O}\left(1/Er\right)\right)$ at large $r$. Similarly, we denote the ingoing solution by $\varphi_{-}$, which has the large-$r$ behavior $\varphi_{-}\left(r\right)=\e^{-iEr}\left(1+\mathcal{O}\left(1/Er\right)\right)$. Their Wronskian $W$ is constant,
\begin{equation}
\label{eq:Wronskian}
W\left[\varphi_{-},\varphi_{+}\right]=\varphi_{-}\varphi_{+}^{\prime}-\varphi_{-}^{\prime}\varphi_{+}=2iE.
\end{equation}
We then define the envelope function $\mathcal{F}\left(r\right)$ as the dressing of the wave by the black-hole geometry relative to flat space,
\begin{equation}
\label{eq:EnvelopeDef}
\phi\left(t,r\right)=\mathcal{F}\left(r\right)\phi_{\text{free}}\left(t,r\right).
\end{equation}
As $r\to\infty$, $\mathcal{F}$ is the transmission amplitude through the collapse geometry, the ratio of the full outgoing wave to the free one on future null infinity. It therefore carries the same phase as $1+i\tilde{A}$ in its $\eta^{-1}$ row of \autoref{eq:eikExp}.

Pulling the carrier $\phi_{\text{free}}$ through the Kerr--Schild derivative $\left(\partial_{t}-\partial_{r}\right)$ turns \autoref{eq:CollapseEOM} into an equation for the envelope alone,
\begin{equation}
\label{eq:EnvelopeEOM}
\mathcal{F}^{\prime\prime}\left(r\right)+2\frac{\varphi_{+}^{\prime}\left(r\right)}{\varphi_{+}\left(r\right)}\mathcal{F}^{\prime}\left(r\right)=\mathcal{D}\left(h\left(r\right)\mathcal{D}\mathcal{F}\right)+\mathcal{C}_{h}\mathcal{F},
\end{equation}
where $\mathcal{D}$ is the Kerr--Schild derivative as it acts on the envelope,
\begin{equation}
\label{eq:D}
\mathcal{D}\mathcal{F}\equiv\left(\partial_{r}+iE+\frac{\varphi_{+}^{\prime}\left(r\right)}{\varphi_{+}\left(r\right)}\right)\mathcal{F},
\end{equation}
and $\mathcal{C}_{h}$ collects the terms independent of $E$,
\begin{equation}
\label{eq:Ch}
\mathcal{C}_{h}\left(r\right)\equiv-\frac{a\left(a-1\right)}{r^{2}}h\left(r\right)-\frac{a}{r}h^{\prime}\left(r\right).
\end{equation}
Treating the right-hand side as a source, the Green's function is
\begin{equation}
\label{eq:GreensFunction}
\mathcal{G}\left(r,r^{\prime}\right)=\frac{1}{2iE}\left[\varphi_{+}\left(r^{\prime}\right)\varphi_{-}\left(r^{\prime}\right)\Theta\left(r-r^{\prime}\right)+\varphi_{+}^{2}\left(r^{\prime}\right)\frac{\varphi_{-}\left(r\right)}{\varphi_{+}\left(r\right)}\Theta\left(r^{\prime}-r\right)\right].
\end{equation}
The first term propagates forward from the source point; the second is the wave reflected back from it. Denoting
\begin{equation}
\label{eq:GBracket}
\mathcal{G}\left[g\right]\left(r\right)\equiv\int_{-v/2}^{\infty}\mathcal{G}\left(r,r^{\prime}\right)g\left(r^{\prime}\right)\dd{r^{\prime}},
\end{equation}
we solve \autoref{eq:EnvelopeEOM} as a Born series,
\begin{equation}
\label{eq:BornSeries}
\mathcal{F}_{n}=\mathcal{G}\left[\mathcal{D}\left(h\mathcal{D}\mathcal{F}_{n-1}\right)\right]+\mathcal{G}\left[\mathcal{C}_{h}\mathcal{F}_{n-1}\right],\qquad\mathcal{F}_{0}=1,
\end{equation}
which can be represented diagrammatically as
\begin{equation}
\label{eq:Born comb}
\mathcal{F}\left(\infty\right) = \sum_{n}\;
\begin{gathered}
    \vspace{-0.5cm}
    \begin{tikzpicture}
        \filldraw[OliveGreen] (-1.575,-0.075) rectangle (-1.425,0.075);
        \node[anchor=north] at (-1.5,-0.15) {\footnotesize $\mathcal{F}_0 = 1$};
        \draw[line width=0.3mm] (-1.5, 0) -- (-0.15, 0);
        \draw[line width=0.3mm] (0.15, 0) -- (1.5, 0);
        \draw[line width=0.3mm] (0, 0) node[cross=0.15cm, RoyalBlue] {};
        \draw[line width=0.3mm, RoyalBlue] (0, 0) circle (0.15cm);
        \node[anchor=south] at (0,0.2) {\footnotesize $\mathcal{V}\left(r_1\right)$};
        \node[anchor=north] at (0.825,-0.1) {\footnotesize $\mathcal{G}\left(r_2, r_1\right)$};
        \node at (1.775, 0) {$\dots$};
        \draw[line width=0.3mm] (2, 0) -- (3.35, 0);
        \draw[line width=0.3mm] (3.65, 0) -- (5, 0);
        \draw[line width=0.3mm] (3.5, 0) node[cross=0.15cm, RoyalBlue] {};
        \draw[line width=0.3mm, RoyalBlue] (3.5, 0) circle (0.15cm);
        \node[anchor=south] at (3.5,0.2) {\footnotesize $\mathcal{V}\left(r_n\right)$};
        \node[anchor=north] at (4.3,-0.1) {\footnotesize $\mathcal{G}\left(\infty, r_n\right)$};
        \filldraw (5, 0) circle (0.075cm);
        \node[anchor=west] at (5.1, 0) {\footnotesize $r \to \infty$};
    \end{tikzpicture}
\end{gathered}
\end{equation}
The square is the free envelope entering at the shell, each cross is the vertex $\mathcal{V}\equiv\mathcal{D}h\mathcal{D}+\mathcal{C}_{h}$ acting on what arrives from the left, each line is the Green's function, and the dot is evaluation at infinity; this is the comb of \autoref{eq:Resummation} at fixed $E$ in position space.

By \autoref{eq:eikExp}, the phase $\chi$ we wish to extract is the $\eta^{-1}$ part of $\log\mathcal{F}\left(\infty\right)$, not of $\mathcal{F}$ itself. Computing $\mathcal{F}$ and then taking its logarithm would generate terms of order $\eta^{-n}$ at order $G^{n}$ whose fate is to cancel. This is already visible in \autoref{s:Matching to the Eikonal Exponent}: at one loop the term $i\tilde{A}_{1}^{\left(-1\right)}$ is canceled in the logarithm by $\frac{1}{2}\left(i\tilde{A}_{0}^{\left(0\right)}\right)^{2}$, leaving only the combination \autoref{eq:chi1amps} at order $\eta^{-1}$. We instead compute the logarithmic derivative, in which such terms never appear. Writing
\begin{equation}
\label{eq:sDef}
s\left(r\right)\equiv\partial_{r}\log\mathcal{F}\left(r\right)=\frac{\mathcal{F}^{\prime}\left(r\right)}{\mathcal{F}\left(r\right)},
\end{equation}
we have
\begin{equation}
\label{eq:LogFInfinity}
\log\mathcal{F}\left(\infty\right)=\log\mathcal{F}\left(-v/2\right)+\int_{-v/2}^{\infty}s\left(r\right)\dd{r},
\end{equation}
so we need the derivative of the Born series \autoref{eq:BornSeries}. This follows from two properties of the Green's function. First, differentiating \autoref{eq:GBracket} with \autoref{eq:GreensFunction}, the local terms of the two branches cancel because $\mathcal{G}$ is continuous at $r=r^{\prime}$, and $\partial_{r}\left(\varphi_{-}/\varphi_{+}\right)=-W/\varphi_{+}^{2}=-2iE/\varphi_{+}^{2}$. This gives
\begin{equation}
\label{eq:PropFactOne}
\partial_{r}\mathcal{G}\left[g\right]\left(r\right)=-\int_{r}^{\infty}K\left(r,r^{\prime}\right)g\left(r^{\prime}\right)\dd{r^{\prime}},
\end{equation}
where
\begin{equation}
\label{eq:KDef}
K\left(r,r^{\prime}\right)\equiv\left(\frac{\varphi_{+}\left(r^{\prime}\right)}{\varphi_{+}\left(r\right)}\right)^{2}
\end{equation}
is the kernel of the reflected branch. Second, from
\begin{equation}
\label{eq:Kderivative}
\partial_{r^{\prime}}K\left(r,r^{\prime}\right)=2\frac{\varphi_{+}^{\prime}\left(r^{\prime}\right)}{\varphi_{+}\left(r^{\prime}\right)}K\left(r,r^{\prime}\right)=2\left(iE+\zeta\left(r^{\prime}\right)\right)K\left(r,r^{\prime}\right),
\end{equation}
where
\begin{equation}
\label{eq:ZetaDef}
\zeta\left(r\right)\equiv\frac{\varphi_{+}^{\prime}\left(r\right)}{\varphi_{+}\left(r\right)}-iE,
\end{equation}
and the definition of $\mathcal{D}$ in \autoref{eq:D}, the operator $\mathcal{D}$ acting under the reflected kernel is a total derivative up to a mismatch term,
\begin{equation}
\label{eq:TotalDerivative}
K\left(r,r^{\prime}\right)\,\mathcal{D}g\left(r^{\prime}\right)=\partial_{r^{\prime}}\!\left[K\left(r,r^{\prime}\right)g\left(r^{\prime}\right)\right]-\zeta\left(r^{\prime}\right)K\left(r,r^{\prime}\right)g\left(r^{\prime}\right).
\end{equation}
Integrating \autoref{eq:TotalDerivative} from $r$ to infinity, the lower boundary term is $-g\left(r\right)$ since $K\left(r,r\right)=1$, and the upper one, $K\left(r,r^{\prime}\right)g\left(r^{\prime}\right)$ as $r^{\prime}\to\infty$, vanishes for $E\to E+i0$, the same prescription that selects the outgoing solution:
\begin{equation}
\label{eq:PropFactTwo}
\int_{r}^{\infty}K\left(r,r^{\prime}\right)\mathcal{D}g\left(r^{\prime}\right)\dd{r^{\prime}}=-g\left(r\right)-\int_{r}^{\infty}\zeta\left(r^{\prime}\right)K\left(r,r^{\prime}\right)g\left(r^{\prime}\right)\dd{r^{\prime}}.
\end{equation}
Applying \autoref{eq:PropFactOne} to \autoref{eq:BornSeries} and \autoref{eq:PropFactTwo} to its first term,
\begin{equation}
\label{eq:FPrime}
\mathcal{F}^{\prime}\left(r\right)=h\left(r\right)\mathcal{D}\mathcal{F}\left(r\right)+\mathcal{L}\mathcal{F}\left(r\right),
\end{equation}
where
\begin{equation}
\label{eq:LDef}
\mathcal{L}\mathcal{F}\left(r\right)\equiv\int_{r}^{\infty}K\left(r,r^{\prime}\right)\left(\zeta h\,\mathcal{D}\mathcal{F}-\mathcal{C}_{h}\mathcal{F}\right)\left(r^{\prime}\right)\dd{r^{\prime}}.
\end{equation}
Dividing \autoref{eq:FPrime} by $\mathcal{F}$ and using \autoref{eq:D} in the form
\begin{equation}
\label{eq:DoverF}
\frac{\mathcal{D}\mathcal{F}}{\mathcal{F}}=s+2iE+\zeta
\end{equation}
gives a closed equation for $s$,
\begin{equation}
\label{eq:sEquation}
\left(1-h\right)s=h\left(2iE+\zeta\right)+\frac{\mathcal{L}\mathcal{F}}{\mathcal{F}}.
\end{equation}
We now expand in $G$. Each factor of $h$ carries one power of $G$, so $h\left(2iE+\zeta\right)$ is of order $G$, and $s$ starts at order $G$ since $\mathcal{F}_{0}=1$. Reading off the coefficient of $G^{n}$ in \autoref{eq:sEquation} gives a recursion,
\begin{equation}
\label{eq:sRecursion}
\left[s\right]_{G^{n}}-h\left[s\right]_{G^{n-1}}=\delta_{n,1}\,h\left(2iE+\zeta\right)+\left[\frac{\mathcal{L}\mathcal{F}}{\mathcal{F}}\right]_{G^{n}},
\end{equation}
in which the right-hand side involves $s$ only at orders below $n$, because $\mathcal{L}$ carries one explicit $h$ or $\mathcal{C}_{h}$. Solving it,
\begin{equation}
\label{eq:sOrderByOrder}
\left[s\right]_{G^{n}}=h^{n}\left(2iE+\zeta\right)+\sum_{k=0}^{n-1}h^{k}\left[\frac{\mathcal{L}\mathcal{F}}{\mathcal{F}}\right]_{G^{n-k}}
\end{equation}
for $n\geq1$. The leading part, $s=2iEh/\left(1-h\right)$, is the derivative of the exponent of \autoref{eq:Ray Tracing from Amplitudes}; we now show that the remaining terms are of order $\eta$ at every order in $G$.

\subsection{\texorpdfstring{Extracting the $\eta^{-1}$ Terms at Every Order in $G$}{Extracting the 1/eta Terms at Every Order in G}}
\label{a:Grading}

The orders of the ingredients are as follows: $h$, $h^{\prime}$ and $\mathcal{C}_{h}$ contain no $E$ and are of order $\eta^{0}$, while $\zeta=\mathcal{O}\left(1/Er^{2}\right)$ is of order $\eta$.

The remainder $\left[\mathcal{L}\mathcal{F}/\mathcal{F}\right]_{G^{n}}$ is of order $\eta$ for every $n\geq1$, so that $\left[s\right]_{G^{n}}$ is of order $\eta^{-1}$. We argue by induction on $n$, assuming that $\left[s\right]_{G^{k}}$ and its derivatives are of order $\eta^{-1}$ for all $k<n$; this is empty at $n=1$, since $s$ starts at order $G$. By \autoref{eq:sRecursion}, $\left[\mathcal{L}\mathcal{F}/\mathcal{F}\right]_{G^{n}}$ involves $s$ only at these lower orders. By \autoref{eq:DoverF}, the combination $\left(\zeta h\mathcal{D}\mathcal{F}-\mathcal{C}_{h}\mathcal{F}\right)/\mathcal{F}$ in \autoref{eq:LDef} is then of order $\eta^{0}$. However, $K$ is oscillatory: integrating \autoref{eq:Kderivative} from $K\left(r,r\right)=1$ gives
\begin{equation}
\label{eq:KOscillatory}
K\left(r,r^{\prime}\right)=\e^{2iE\left(r^{\prime}-r\right)}\exp\left[2\int_{r}^{r^{\prime}}\zeta\left(\tilde{r}\right)\dd{\tilde{r}}\right],
\end{equation}
a pure oscillation at frequency $2E$ times a factor of order $\eta^{0}$. Furthermore, the order-$\eta^{0}$ statement applies to $\mathcal{D}\mathcal{F}\left(r^{\prime}\right)/\mathcal{F}\left(r^{\prime}\right)$ inside the integral, whereas \autoref{eq:LDef} is divided by $\mathcal{F}\left(r\right)$, so we pick up an additional factor
\begin{equation}
\label{eq:FRatio}
\frac{\mathcal{F}\left(r^{\prime}\right)}{\mathcal{F}\left(r\right)}=\exp\int_{r}^{r^{\prime}}s\left(\tilde{r}\right)\dd{\tilde{r}},
\end{equation}
which is not small: at order $G^{n}$ its expansion contains powers of $\int s$ up to the $\left(n-1\right)$th, of order $\eta^{-\left(n-1\right)}$. The integrand is therefore not graded; the integral is. Observing that
\begin{equation}
\label{eq:ExpDerivative}
\e^{2iE\left(r^{\prime}-r\right)}=\partial_{r^{\prime}}\left[\e^{2iE\left(r^{\prime}-r\right)}/2iE\right]
\end{equation}
and integrating by parts, the upper limit removed by $E\to E+i0$, we find
\begin{equation}
\label{eq:EndpointIBP}
\int_{r}^{\infty}\e^{2iE\left(r^{\prime}-r\right)}g\left(r^{\prime}\right)\dd{r^{\prime}}=-\frac{g\left(r\right)}{2iE}-\frac{1}{2iE}\int_{r}^{\infty}\e^{2iE\left(r^{\prime}-r\right)}g^{\prime}\left(r^{\prime}\right)\dd{r^{\prime}}
\end{equation}
for any $g$ decaying at infinity. In our case
\begin{equation}
\label{eq:gDef}
g\left(r^{\prime}\right)=\exp\left[\int_{r}^{r^{\prime}}\left(s+2\zeta\right)\left(\tilde{r}\right)\dd{\tilde{r}}\right]\left(\zeta h\left(s+2iE+\zeta\right)-\mathcal{C}_{h}\right)\left(r^{\prime}\right).
\end{equation}
The boundary term in \autoref{eq:EndpointIBP} is $g$ at the endpoint $r^{\prime}=r$, where the exponential in \autoref{eq:gDef} equals one, leaving the order-$\eta^{0}$ combination multiplied by $1/2iE$: order $\eta$. Each further integration by parts brings one more factor of $1/2iE$ and one more derivative of $g$ at the endpoint. A derivative acting on the exponential produces a factor $s+2\zeta$, of order $\eta^{-1}$ by the hypothesis, which the $1/2iE$ pays for exactly; acting elsewhere it does not lower the order, since a derivative in $r$ never does. So every term of the resulting series is of order $\eta$. Only finitely many terms reach that order: the combination in \autoref{eq:gDef} carries one $h$ and each $s$ another, so at order $G^{n}$ at most $n-1$ factors of $s$ can appear, and after $n+1$ integrations by parts the remainder is of order $\eta^{2}$ or smaller. Hence $\left[\mathcal{L}\mathcal{F}/\mathcal{F}\right]_{G^{n}}$ is of order $\eta$, and reading off \autoref{eq:sOrderByOrder}, $\left[s\right]_{G^{n}}=h^{n}\left(2iE+\zeta\right)+\mathcal{O}\left(\eta\right)$ is of order $\eta^{-1}$, together with its derivatives since the endpoint series is local, which completes the induction.

The shell term in \autoref{eq:LogFInfinity} is of order $\eta^{0}$. At the shell only the reflected branch of \autoref{eq:GreensFunction} is present, and by \autoref{eq:PropFactOne} it equals
\begin{equation}
\label{eq:Kick}
\mathcal{F}\left(-\frac{v}{2}\right)-1=\left.-\frac{\varphi_{+}\left(r\right)\varphi_{-}\left(r\right)\mathcal{F}^{\prime}\left(r\right)}{2iE}\right|_{r=-v/2}.
\end{equation}
Since $\mathcal{F}^{\prime}/\mathcal{F}=s$ is of order $\eta^{-1}$ by the result above, $\mathcal{F}\left(-v/2\right)-1$ is of order $\eta^{0}$ relative to $\mathcal{F}\left(-v/2\right)$, and so is $\log\mathcal{F}\left(-v/2\right)$.

With these two results, the sum in \autoref{eq:sOrderByOrder} is of order $\eta$, and so is $h^{n}\zeta$; hence
\begin{equation}
\label{eq:sLeading}
\left[s\right]_{G^{n}}=2iE\,h^{n}\left(r\right)+\mathcal{O}\left(\eta\right).
\end{equation}
Inserting this into \autoref{eq:LogFInfinity}, the shell term contributing nothing at this order,
\begin{equation}
\label{eq:PhaseAtOrderN}
\left[\log\mathcal{F}\left(\infty\right)\right]_{\eta^{-1},G^{n}}=2iE\int_{-v/2}^{\infty}h^{n}\left(r\right)\dd{r}
\end{equation}
for every $n\geq1$. By \autoref{eq:eikExp}, the left-hand side is $i\chi_{n-1}$. Since \autoref{eq:sLeading} has no term of order $\eta^{-k}$ with $k\geq2$ at any order in $G$, neither has $\log\mathcal{F}\left(\infty\right)$, so the exponentiation of \autoref{eq:eikExp} holds with $\Delta=\mathcal{O}\left(\eta^{0}\right)$ at every order. Summing \autoref{eq:PhaseAtOrderN} over $n$,
\begin{equation}
\label{eq:PhaseSummed}
i\chi=2iE\sum_{n\geq1}\int_{-v/2}^{\infty}h^{n}\left(r\right)\dd{r}=2iE\int_{-v/2}^{\infty}\frac{h\left(r\right)}{1-h\left(r\right)}\dd{r},
\end{equation}
the exponent of \autoref{eq:Ray Tracing from Amplitudes}. 

\subsection{Comparison with the Leading Eikonal}
\label{a:ComparisonLeadingEikonal}

The leading-eikonal evaluation of \autoref{s:The Case of Reversed Limits} keeps, at order $G^{n}$, only the term of order $\eta^{-n}$. In the present language it truncates both the vertex and the propagator of \autoref{eq:BornSeries}: every $\mathcal{D}$ is replaced by $2iE$ and $\mathcal{C}_{h}$ is dropped, and \autoref{eq:GreensFunction} is replaced by its forward branch with $\varphi_{+}\varphi_{-}$ set to one, leaving only $\Theta\left(r-r^{\prime}\right)/2iE$. The truncated recursion is
\begin{equation}
\label{eq:TruncatedRecursion}
\mathcal{F}_{n}\left(r\right)=2iE\int_{-v/2}^{r}h\left(r^{\prime}\right)\mathcal{F}_{n-1}\left(r^{\prime}\right)\dd{r^{\prime}},
\end{equation}
whose solution 
\begin{equation}
    \mathcal{F}_{n}=\frac{\left(i\chi_{0}\right)^{n}}{n!}
\end{equation}
is the tree-level exponentiation of \autoref{s:The Case of Reversed Limits}. Its logarithm is $i\chi_{0}$ alone: the two truncations together remove the term $h\,\mathcal{F}^{\prime}$ of \autoref{eq:FPrime}, which is what generates every $h^{n}$ with $n\geq2$ in \autoref{eq:sEquation}. 

\section{Additional Examples}
\label{a:AdditionalExamples}
In this subsection, we include additional evidence for the universal Stokes line in the plane of outgoing energy $E$ upon using exponential coordinates. The examples we consider are an observer far away from a black hole surrounded by a Hernquist dark matter cloud, as studied for example in \cite{Heidari:2026swh}, and a static observer a finite distance away from a 4d Schwarzschild black hole, as studied for example in \cite{Barbado:2011dx}.

\subsection{Schwarzschild--Hernquist}

The Schwarzschild--Hernquist solution describes a black hole immersed in a dark matter halo with density profile
\begin{equation}
    \rho(r) = \rho_0 \frac{R}{r} \left( 1 + \frac{r}{R} \right)^{-3}
\end{equation}
where $R$ and $\rho_0$ are the characteristic radius and density of the halo. The relevant Kerr--Schild potential at linear order is
\begin{equation}
    h(r) = \frac{r_s}{r} + \frac{r_{\text{DM}}}{r+R},
\end{equation}
where $r_{\text{DM}}$ encodes the mass density of the dark matter cloud.

From \autoref{eq:Ray Tracing from Amplitudes}, we can compute the ray tracing relationship and find 
\begin{equation}
    u\left(v\right) = v - \frac{2 r_+\left(r_+ + R\right)}{r_+ - r_-}\log\left(\frac{v + 2 r_+}{-\ell}\right) + \frac{2 r_-\left(r_- + R\right)}{r_+ - r_-} \log\left(\frac{v + 2 r_-}{-\ell}\right),
\end{equation}
where the horizon locations are given by
\begin{equation}
r_{\pm} = \frac{1}{2}\left(r_s + r_{\text{DM}} - R \pm \sqrt{\left(R - r_s - r_{\text{DM}}\right)^2 + 4 r_s R}\right).
\end{equation}
We can then express the second generalized Hawking amplitude, $\mathcal{B}$ in exponential coordinates, using
\begin{equation}
\operatorname{e}^z = \frac{v + 2r_+}{-\ell},
\end{equation}
as before, yielding 
\begin{equation}
\label{eq:General Form B}
    \mathcal{B} \propto \int_{-\infty}^{\infty} \dd{z} \e^{\tilde{S}\left(z\right)},
\end{equation}
where the action is
\begin{equation}
\begin{aligned}
    \tilde{S}\left(z\right) &= \left(1 + \frac{2i E r_+\left(r_+ + R\right)}{r_+ - r_-}\right) z + i\ell\left(E_0 + E\right) \e^z \\
    &\quad - \frac{2i E r_-\left(r_- + R\right)}{r_+ - r_-}\log\left(\e^z - \frac{2 \left(r_+ - r_-\right)}{\ell}\right)
\end{aligned}
\end{equation}
and the proportionality factor is irrelevant for our purposes.

In order to retrieve the Stokes line corresponding to the Hawking temperature, we may compute the condition $\operatorname{Im}[S(z +2\pi i )] = \operatorname{Im} [S(z)]$, which produces
\begin{equation}
    \text{Im}[E] = \frac{r_+ - r_-}{2r_+ (r_+ +R)}. 
\end{equation}
The Hawking temperature is then, at this order in the dark matter profile density, 
\begin{equation}
    T_H = \frac{r_+ - r_-}{4\pi r_+ \left(r_+ + R\right)},
\end{equation}
corresponding to the solution found in \cite{Heidari:2026swh}. Upon computing $\mathcal{B}$ in the saddle point approximation, one will then find a Bose--Einstein spectrum with the correct temperature. 

\subsection{Static Observer}
Another case that exhibits unique Stokes phenomena is that of a static observer sitting at a fixed finite radius in a Schwarzschild background. As computed in \cite{Barbado:2011dx}, such an observer is associated with a ray-tracing relation of
\begin{align}
    u_{\text{FR}}(v) = \left( 1- \frac{r_s}{R}\right)^{1/2} &\biggl[-\frac{4r_s}{3} - 2r_s\biggl[ \log \left( \left( \frac{3}{2r_s} \left( -2r_s - v \right) +1 \right)^{1/3} -1\right) \\ &+ \sum_{n=1}^3 \frac{1}{n} \left( \frac{3}{2r_s} \left(-2r_s - v \right) +1 \right)^{n/3} \biggr] \biggr],\nonumber
\end{align}
where $R$ is the radius of the fixed-radius observer, and $v_0$ is the location of the horizon, in analogy to the standard asymptotic observer. Since the observer is no longer on future null-infinity, however, it is not clear that our derivation from the one-to-one transition amplitude follows through cleanly, particularly since the use of the $S$-matrix assumes asymptotically free states. It would be interesting to derive this result using a scattering amplitudes calculation.

We may once again switch to exponential coordinates by unraveling the logarithm
\begin{equation}
    \operatorname{e}^z = \left(\frac{3}{2r_s}\left(-2r_s - v\right) + 1\right)^{1/3} - 1.
\end{equation}
In doing so, we find that the generalized amplitudes takes the same form as in \autoref{eq:General Form B}, but now with the action 
\begin{equation}
\begin{aligned}
S(z) &= z+  2\log \left( 1+\e^z \right) - iE_0 \left( v_0 -\frac{2r_s}{3}((\e^z +1)^3 -1) \right) \\
    &\quad -iE \left( 1 - \frac{r_s}{R}\right)^{1/2} \Biggl\{ v_0 + \frac{2r_s}{3} - 2r_s \Biggl[ \sum_{n=1}^3 \frac{1}{n} (\e^z +1)^n + z \Biggr] \Biggr\} .
\end{aligned}
\end{equation}
Once again the proportionality factor is not important for our argument. 

To determine the saddle points in the asymptotic $|E|\rightarrow \infty$ region, we may write the saddle point equation, after a variable change $e^z +1 \mapsto w$, as
\begin{equation}
    w^4 - \frac{E_0}{E_0 + E \sigma} w^3 - \frac{3i}{2E_0 r_s + 2 E r_s \sigma} w + \frac{i}{E_0 r_s +E r_s \sigma} = 0\, ,
\end{equation}
where
\begin{equation}
    \sigma = \sqrt{1 - \frac{r_s}{R}}.
\end{equation}
For asymptotically large $E$, this can be solved approximately as
\begin{align}
    w_{\ast,k} &= \left( \frac{-i}{E_0 r_s +E r_s \sigma} \right)^{1/4} = \frac{\e^{i(\frac{k\pi}{2} - \frac{\pi}{8})}}{(E_0 r_s +E r_s \sigma)^{1/4}}, \quad k= 0,1,2,3. \\
    z_{*,(n,k)} &= \log (w_* - 1)+2\pi i n = \log \left(\frac{\e^{i(\frac{k\pi}{2} - \frac{\pi}{8})}}{(E_0 r_s +E r_s \sigma)^{1/4}} -1\right) + 2\pi i n , \quad n\in \mathbb{Z}.
\end{align}
As seen previously, we now use these saddle points to compute the necessary but insufficient condition for a Stokes line; that $\text{Im}[S(z_{\ast,(i,j)})] = \text{Im}[S(z_{\ast,(k,\ell )})]$. Although we could compute this condition for any pair of saddles, numerical analysis rules out all but two pairs: $(z_{*(0,0)}, z_{*(1,0)})$ and $(z_{*(0,0)}, z_{*(0,1)})$. Substituting these into the exponent $S(z)$, dropping all subdominant terms in the asymptotic $E$ expansion, and solving the $\text{Im}[S(z_{*,(i,j)})] = \text{Im}[S(z_{*,(k,\ell )})]$ condition for each pair gives:
\begin{subequations}
\begin{align}
    (z_{*(0,0)}, z_{*(1,0)}): \quad \text{Im}[E] &= \frac{1}{2r_s \sigma}, \\
    (z_{*(0,0)}, z_{*(0,1)}): \quad \text{Im}[E] &= \frac{3}{4r_s \sigma}.
\end{align}
\end{subequations}
The temperature Stokes line is the one which corresponds to a jump from a finite to an infinite number of relevant saddles, which turns to be the first of these two lines. This gives a temperature
\begin{equation}
    T_H = \frac{1}{4\pi \sigma r_s},
\end{equation}
which is a blue-shifted measure of the temperature, as expected and agrees with \cite{Barbado:2011dx}.

\newpage

\bibliographystyle{JHEP}
\bibliography{Bibliography.bib}

\end{document}